\documentclass[
superscriptaddress,
amsmath, amssymb,
aps,
prl,
dblfloatfix,
nobalancelastpage,
reprint
]{revtex4-1}

\usepackage{graphicx}
\usepackage{dcolumn}
\usepackage{bm}
\usepackage{hyperref}
\usepackage{caption}
\usepackage{amsmath}
\usepackage[subrefformat=parens,labelformat=parens]{subfig}
\usepackage{placeins}
\usepackage{floatrow}
\usepackage{float}

\begin{document}

\preprint{APS/123-QED}

\title{Spin-Polarized Fractional Corner Charges and Their Photonic Realization}

\author{Ran Gladstein Gladstone}
\affiliation{School of Applied and Engineering Physics, Cornell University, Ithaca, NY 14853}
\email{rag298@cornell.edu}
\author{Minwoo Jung}
\affiliation{Department of Physics, Cornell University, Ithaca, NY 14853}
\author{Gennady Shvets}
\affiliation{School of Applied and Engineering Physics, Cornell University, Ithaca, NY 14853}

\date{\today}

\begin{abstract}
We demonstrate that a spin degree of freedom can introduce additional texture to higher order topological insulators (HOTIs), manifesting in novel topological invariants and phase transitions. Spin-polarized mid-gap corner states of various multiplicities are predicted for different HOTI phases, and novel bulk-boundary correspondence principles are defined based on bulk invariants such as total and spin corner charge. Those are shown to be robust to spin-flipping perturbations. Photonic realizations of spin-linked topological phases are demonstrated in engineered systems using pseudo-spin.
\end{abstract}

\maketitle

The interplay between spin and topology has long been of interest to physicists~\cite{kane2005quantum,bernevig2006quantum,jungwirth2012spin,kimura2007room,qi2006topological}. Shortly after the prediction of time-reversal (TR) preserving Chern insulators~\cite{haldane1988model}, the concept was extended to spinful electrons, leading to the prediction~\cite{bernevig2006Science} and experimental realization of the quantum spin Hall (QSH) effect~\cite{khang2018conductive,jamali2015giant}. The extension to spinful systems typically involves doubling of the Hilbert space to include two similar copies of the original spinless system~\cite{bernevig2006Science,asboth2016short}, and can have significant effect on the topology of the system. For example, while the Chern insulator has a $\mathbb{Z}$ topological invariant, a spin Chern insulator possesses a $\mathbb{Z}_2$ topological invariant due to TR and inversion symmetries~\cite{kitaev2009periodic}. The addition of spin to the system has, therefore, changed its topological classification.

More recently, these concepts have been extended to higher order topological insulators (HOTIs)~\cite{benalcazar2017quantized, schindler2018higher, peterson2018quantized,franca2018anomalous,schindler2019fractional,kooi2021bulk,freeney2020edge,kempkes2019robust}, and their photonic/phononic counterparts (PHOTIs)~\cite{xie2018second,jung2020nanopolaritonic,xie2019visualization,chen2019direct, chen2021dual, noh2018topological}. A HOTI is a $d$-dimensional topological insulator with a bulk topological invariant predicting topological states localized on some of its $d-n$ dimensional terminations, where $n\in \mathbb{N}$ and $2 \leq n \leq d$. It is natural to ask if spinful HOTIs have any emerging topological properties and classification that distinguish them from their spinless counterparts. Most of research of spinful HOTIs has so far concentrated on the systems with fermionic TR symmetry~\cite{kooi2021bulk,schindler2019fractional} or time-dependent drive~\cite{franca2018anomalous}.

In this Letter we examine the effects of adding the spin degree of freedom (DOF) to HOTIs when fermionic TR symmetry is broken yet bosonic TR symmetry is maintained~\cite{maczewsky2020fermionic,asboth2016short}. Our analysis also applies to pseudo-spin DOFs, such as layer~\cite{qiao2011two,park2019higher,pesin2012spintronics,lu2018valley} and polarization~\cite{ma2015guiding,ma2017scattering,gladstone2019photonic}. We use the latter to emulate spinful PHOTIs based on engineered microwave structures. The specific 2D lattice models and their corresponding periodic electromagnetic waveguides considered in this Letter possess $C_6$ symmetry, and are additionally endowed with: Kekulé texture~\cite{wu2016topological,jung2020nanopolaritonic}, spin DOF, and a symmetry-preserving spin-flipping perturbation coupling the spin-$\uparrow$ and spin-$\downarrow$ components. The latter will be assumed in the form of Kane-Mele spin-orbit coupling (SOC)~\cite{kane2005quantum}. For spinless models, such topological crystalline insulators (TCIs) have been shown to support a HOTI phase with a quantized corner charge of $Q_c = 1/2$~\cite{benalcazar2019quantization}.

Below we demonstrate that when an independent spin subspace is introduced, two non-trivial phases can emerge: a phase possessing two quantized spin-polarized corner charges $Q_c^{\uparrow,\downarrow} = 1/2$ (spin-HOTI), and a phase characterized by a single quantized corner charge $Q_c=1/2$ (HOTI). While in the absence of SOC the spin-HOTI phase can be thought of as being independently topological for each of the spin states~\cite{chen2021dual}, such simple interpretation is no longer valid in the presence of a finite spin-flipping SOC. Which one out of these two phases is realized depends on the strength of the SOC that couples the two spin states, and on the individual topological properties of each spin component in the absence of such coupling. We further demonstrate that the distinct properties of the spin-HOTI and HOTI phases manifest themselves in different multiplicities and spin textures of their corresponding zero-energy corner states. When effective fermionic TR symmetry is restored~\cite{wu2016topological,szameit2020fermionic}, the corner states delocalize and merge with topologically-protected edge states.

\begin{figure}
    \centering
    \includegraphics[width=\columnwidth]{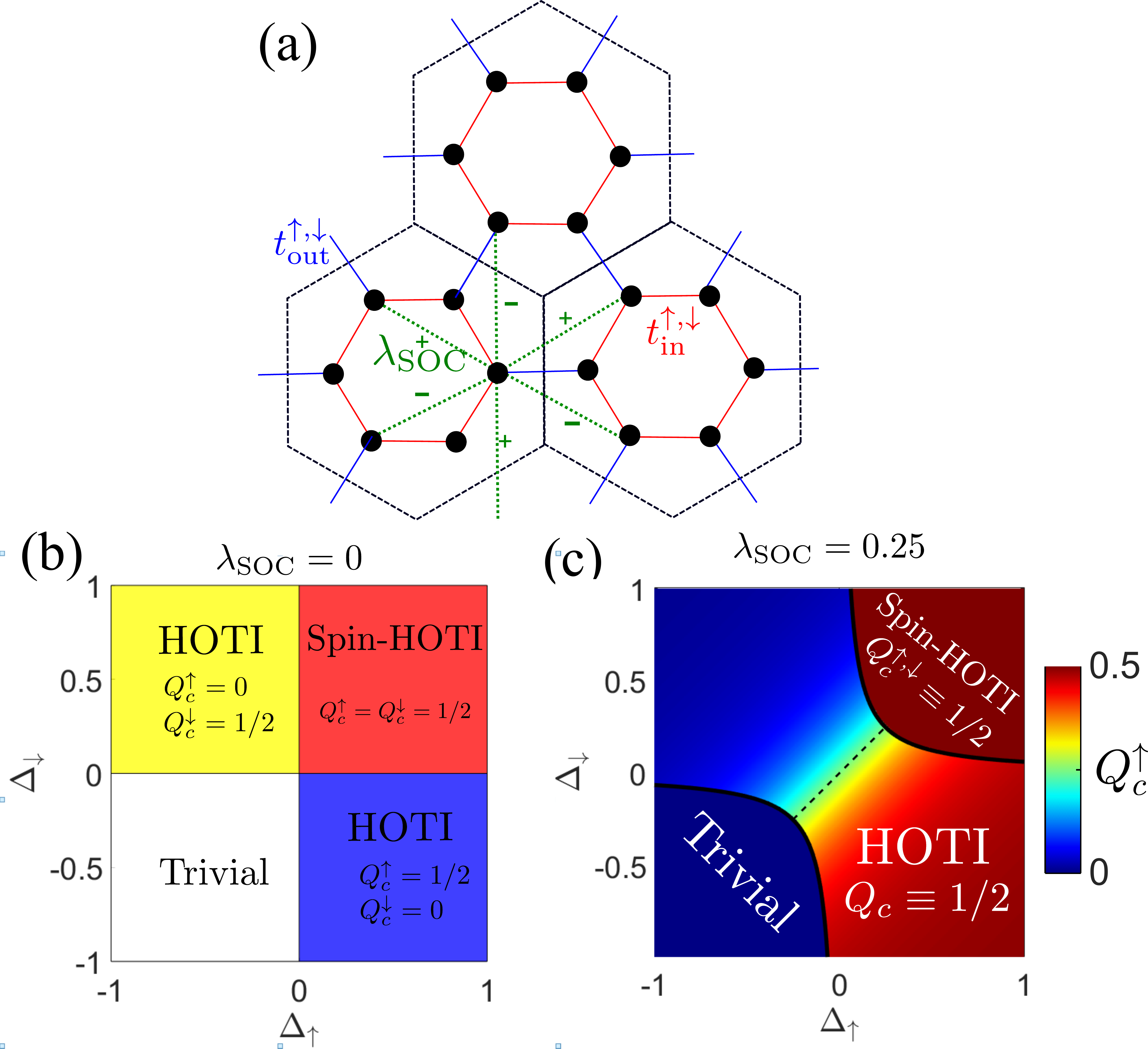}
    \caption{Spinful HOTI: the tight-binding (TB) model and its phase diagrams. \textbf{(a)} Schematic of the TB model with definitions of intra-cell (red lines) and inter-cell (blue lines) hopping amplitudes. Black circles: lattice sites, dashed hexagons: unit cells, green dashed lines: spin-orbit coupling (SOC) amplitudes $\lambda_{\text{SOC}}$ and their signs ($\pm$).\textbf{(b,c)} Phase diagram at constant $\lambda_\text{SOC}=0$ (b) and $\lambda_\text{SOC}=0.25$ (c). Solid curves: bulk bandgap closure at  $\lambda_\text{SOC} = \lambda_{\text{SOC}}^{\rm th}(\Delta_\uparrow, \Delta_\downarrow)$, color bar: $Q_c^{\uparrow}$ for trivial, spin-topological, and topological (not quantized) phases. Dashed line: spin-degeneracy $\Delta_\uparrow = \Delta_\downarrow$ corresponding to quantum spin-Hall (QSH) phase.}
    \label{fig:phase}
\end{figure}

Our starting point is the tight binding (TB) model on a honeycomb lattice schematically shown in Fig.~\ref{fig:phase}(a):
\begin{equation}\label{eq:TB}
\begin{split}
    H = \sum_{\langle ij \rangle} t_{\text{in}}^{\uparrow} c_{i\uparrow}^{\dagger} c_{j\uparrow} + \sum_{\langle i'j' \rangle}t_{\text{out}}^{\uparrow} c_{i'\uparrow}^{\dagger} c_{j'\uparrow}
    +\sum_{\langle ij \rangle} t_{\text{in}}^{\downarrow} c_{i\downarrow}^{\dagger} c_{j\downarrow} \\
    + \sum_{\langle i'j' \rangle}t_{\text{out}}^{\downarrow} c_{i'\downarrow}^{\dagger} c_{j'\downarrow}
    + \sum_{\langle\langle ij \rangle\rangle \alpha \beta} \frac{i}{3\sqrt{3}} \lambda_{\text{SOC}} \nu_{ij}s^x_{\alpha \beta}c_{i\alpha}^{\dagger} c_{j \beta},
\end{split}
\end{equation}
where the first (third) term describes the nearest neighbor intra unit cell hopping of the spin up (down) electrons, the second (fourth) term describes the nearest neighbor inter unit cell hopping of the spin-$\uparrow$ ($\downarrow$) electrons, and the fifth term describes the next-nearest neighbor Kane-Mele SOC~\cite{kane2005quantum}. Without loss of generality, we assume that all hopping amplitudes are positive, and that the energy scale is defined by setting $\left( t_{\text{out}}^{\uparrow,\downarrow} + t_{\text{in}}^{\uparrow,\downarrow} \right)/2 = 1$.

The difference between inter- and intra-cell hopping amplitudes, $\Delta_{\uparrow, \downarrow} \equiv  t_{\text{out}}^{\uparrow,\downarrow} - t_{\text{in}}^{\uparrow,\downarrow}$, determines the topology of the two uncoupled spin sub-spaces in the absence of SOC: each spin component's subspace behaves as a HOTI for $ \Delta_{\uparrow,\downarrow} > 0$, or a trivial insulator for $\Delta_{\uparrow,\downarrow} < 0$, as shown in Fig.~\ref{fig:phase}(b). The topological phase transitions at $ \Delta_{\uparrow,\downarrow} = 0$ are the consequences of the band inversions between $p$- and $d$-orbital mode profiles~\cite{jung2020nanopolaritonic,wu2016topological}: the energies of the $p$-orbital bands are below (above) the bandgap for trivial (topological) insulators.

The unperturbed spin corner charges $\tilde{Q}_c^{\uparrow,\downarrow}$ are separately defined for each spin subspace~\cite{benalcazar2019quantization} in a manner similar to the spin Chern number~\cite{bernevig2006Science}. Corner charges can be calculated based on the symmetry properties of the propagation bands below the bandgap at the high-symmetry $\Gamma$ and $M$ points of the Brillouin zone~\cite{benalcazar2019quantization}:
\begin{equation}
    \tilde{Q}_c^{\uparrow,\downarrow}  = \left( \#M_{\uparrow,\downarrow} -\#\Gamma_{\uparrow,\downarrow} \right)/4
    \label{eq:Qc_up_down}
\end{equation}
where $\#M_\uparrow$ ($\# \Gamma_\uparrow$) is the number of $C_2$-invariant spin-$\uparrow$ modes below the band gap at the $M$ ($\Gamma$) point (same for spin-$\downarrow$ modes). Under this definition of $\tilde{Q}_c^{\uparrow,\downarrow}$, three distinct cases can be identified: (i) $\tilde{Q}_c^{\uparrow} = \tilde{Q}_c^{\downarrow} = 1/2$ (topological for both spins when $\Delta_{\uparrow,\downarrow} > 0$), (ii) $\left[ \tilde{Q}_c^{\uparrow} = 0, \tilde{Q}_c^{\downarrow} = 1/2 \right]$ or $\left[ \tilde{Q}_c^{\uparrow} = 1/2, \tilde{Q}_c^{\downarrow} = 0 \right]$ (topological for only one spin when $\Delta_{\uparrow} \Delta_{\downarrow} < 0$), and (iii) $\tilde{Q}_c^{\uparrow} = \tilde{Q}_c^{\downarrow} = 0$ (trivial for both spins when $\Delta_{\uparrow,\downarrow} < 0$).

When the spin components are coupled through the SOC, the expressions for $\tilde{Q}_c^{\uparrow,\downarrow}$ are no longer quantized, and must be redefined because of the mixing between the spin-$\uparrow$ and spin-$\downarrow$ contents of the propagation bands. Therefore, we modify the definitions of the corner charges by weighing the contribution of each band by its spin-$\uparrow / \downarrow$ content:
\begin{equation}
    Q_c^{\uparrow,\downarrow} = \frac{1}{4}\sum_i \left( \left|\langle \uparrow,\downarrow |\psi_i(M) \rangle\right|^2 - \left|\langle \uparrow,\downarrow |\psi_i(\Gamma)\rangle\right|^2\rangle \right),\label{eq:Q_prob}
\end{equation}
where $|\psi_i(\mathbf{k})\rangle$ is the wave function of a $C_2$-invariant band $i$ below the band gap at the $\mathbf{k}=M,\Gamma$ points.

It can be shown that $Q_c^{\uparrow,\downarrow}$ defined by Eq.(\ref{eq:Q_prob}) are quantized at $Q_c^{\uparrow,\downarrow}=1/2$ for the spin-HOTI, and at $Q_c^{\uparrow,\downarrow}=0$ for the trivial phase. Briefly, the probability conservation for all the bands below the bandgap ensures that $Q_c^{\uparrow,\downarrow}=\tilde{Q}_c^{\uparrow,\downarrow}$ for these two cases. This is due to the lack of cross-band transitions generated by the $C_6$ symmetry-preserving SOC: the $d$ ($p$) orbitals for both spins are on the same side the bandgap for both spins, and the SOC only connects the orbitals of one spin component to the same orbitals of the other component (see the SOM for details).

On the contrary, $Q_c^{\uparrow,\downarrow}$ corner charges are not separately quantized in the HOTI phase. The cause of this transition from either trivial to HOTI (if $\Delta_{\uparrow \downarrow} <0$) phase, or from spin-HOTI to HOTI (if $\Delta_{\uparrow \downarrow} > 0$) is the band inversion by the strong SOC coupling satisfying $\lambda_{\text{SOC}} > \lambda_{\text{SOC}}^{\rm (th)}$. Here  $\lambda_{\text{SOC}}^{\rm (th)} \equiv \sqrt{\Delta_{\uparrow} \Delta_{\downarrow}}$ is the threshold value of the SOC strength corresponding to the closing of the bulk bandgap, and followed by band inversion. However, the total spin-independent corner charge $Q_c = Q_c^{\uparrow} + Q_c^{\downarrow}=1/2$ remains quantized in the HOTI phase, thus making it topologically nontrivial. Phase diagrams in the $\left( \Delta_{\downarrow}, \Delta_\uparrow \right)$ space showing one trivial and two topological (spin-HOTI and HOTI) phases are shown in Fig.~\ref{fig:phase}(b),(c) for constant $\lambda_\text{SOC}$. The spin-degeneracy region $\Delta_{\uparrow} = \Delta_{\downarrow}$ represented by the dashed line in Fig.~\ref{fig:phase}(c) corresponds to the fermionic TR symmetric QSH phase~\cite{wu2016topological,szameit2020fermionic} embedded in a bosonic TR symmetric HOTI phase (see SOM).

\begin{figure}
    \centering
    \includegraphics[width=\columnwidth]{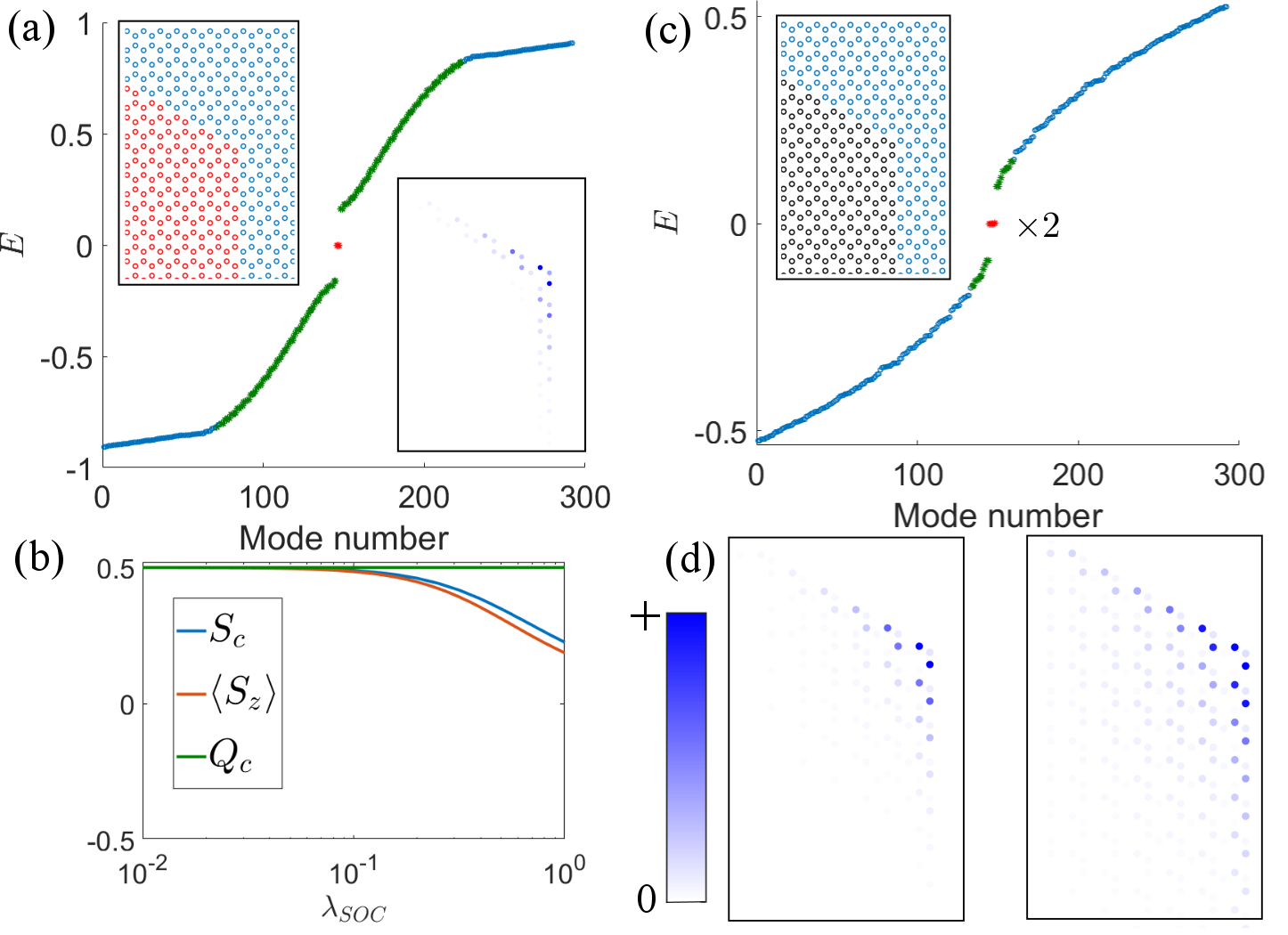}
    \caption{Properties of corner-shaped domain walls between a topological (\textbf{(a)} HOTI and \textbf{(c)} spin-HOTI) and a trivial phase: a TB model. Upper-left inset in \textbf{(a)}: domain wall shape.  Energy spectra: bulk (blue), edge (green), and corner (red) modes for the HOTI/trivial \textbf{(a)} and spin-HOTI/trivial \textbf{(c)} interfaces. Lower-right inset in \textbf{(a)}: intensity of the single ZCS. \textbf{(b)} Comparison of the bulk invariant $Q_c$ and $S_c$ (HOTI phase) and $\langle S_z \rangle$ for the single ZCS in (a). \textbf{(d)} Intensities of the two ZCSs in (c). TB model parameters: see Table S1.}
    \label{fig:TB_corner}
\end{figure}

To investigate the existence and multiplicity of zero-energy corner states (ZCSs), we consider an interface between trivial and topological domains containing a single $120^{\circ}$ corner, as shown in the upper-left inset in Fig.\ref{fig:TB_corner}(a). In the case when the topological phase is a HOTI, its single quantized topological charge $Q_c=1/2$ manifests itself as a single spin-polarized ZCS. The spectrum for such corner-containing domain wall is calculated using a finite-domain TB calculation, and the corresponding spectrum comprised of a continuum of bulk/edge modes and a corner state is shown in Fig.\ref{fig:TB_corner}(a) (see Table S1 for the TB parameters). The existence of the ZCS is a direct consequence of the quantized bulk invariant $Q_c$ and of the chiral symmetry present in a $C_6$ lattice with Kekule distortion, both of which are preserved even with finite SOC (see SOM).

The resulting ZCS has a non-trivial spin texture because the SOC term mixes the two spin components. It can be characterized by the spin imbalance $S_c = Q_c^\uparrow - Q_c^\downarrow \neq 0$. Although the bulk quantity $S_c$ is not quantized in the HOTI phase, our calculation of $\langle S_z \rangle$ from the corner local density of states (LDOS)~\cite{peterson2020fractional} (see SOM) shows that $S_c \approx \langle S_z \rangle$ for a wide range of the SOC strengths $\lambda_{\rm SOC}$, as shown in Fig.\ref{fig:TB_corner}(b). Therefore, the bulk quantity $S_c$ is a useful measure of the experimentally-measurable corner deficit of the spin polarization $\langle S_z \rangle$. 

The existence of the separate quantized spin charges $Q_c^{\uparrow}=Q_c^{\downarrow}=1/2$ in the spin-HOTI phase guarantees a degenerate pair of ZCSs (see Fig.~\ref{fig:TB_corner}(c,d)) with distinct spin textures, as long as $\Delta_{\uparrow,\downarrow} > 0$ and the SOC coupling strength satisfies $\lambda_{\text{SOC}} < \lambda_{\text{SOC}}^{\rm (th)}$. When the latter condition is violated, a phase transition into the HOTI phase takes place, the quantization of the spin charges $Q_c^{\uparrow \downarrow}$ is lost, and the two corner states merge into one ZCS. The localization lengths of the two corner states are generally not equal (see Fig.~\ref{fig:TB_corner}(d)), and are determined by the hopping amplitudes, $\lambda_{\text{SOC}}$, as well as the size of the bandgap between the edge states. The latter can be controlled by "roughening" the corner-adjacent domain edges~\cite{jung2020nanopolaritonic} (see SOM).

\begin{figure}
    \centering
    \includegraphics[width=\columnwidth]{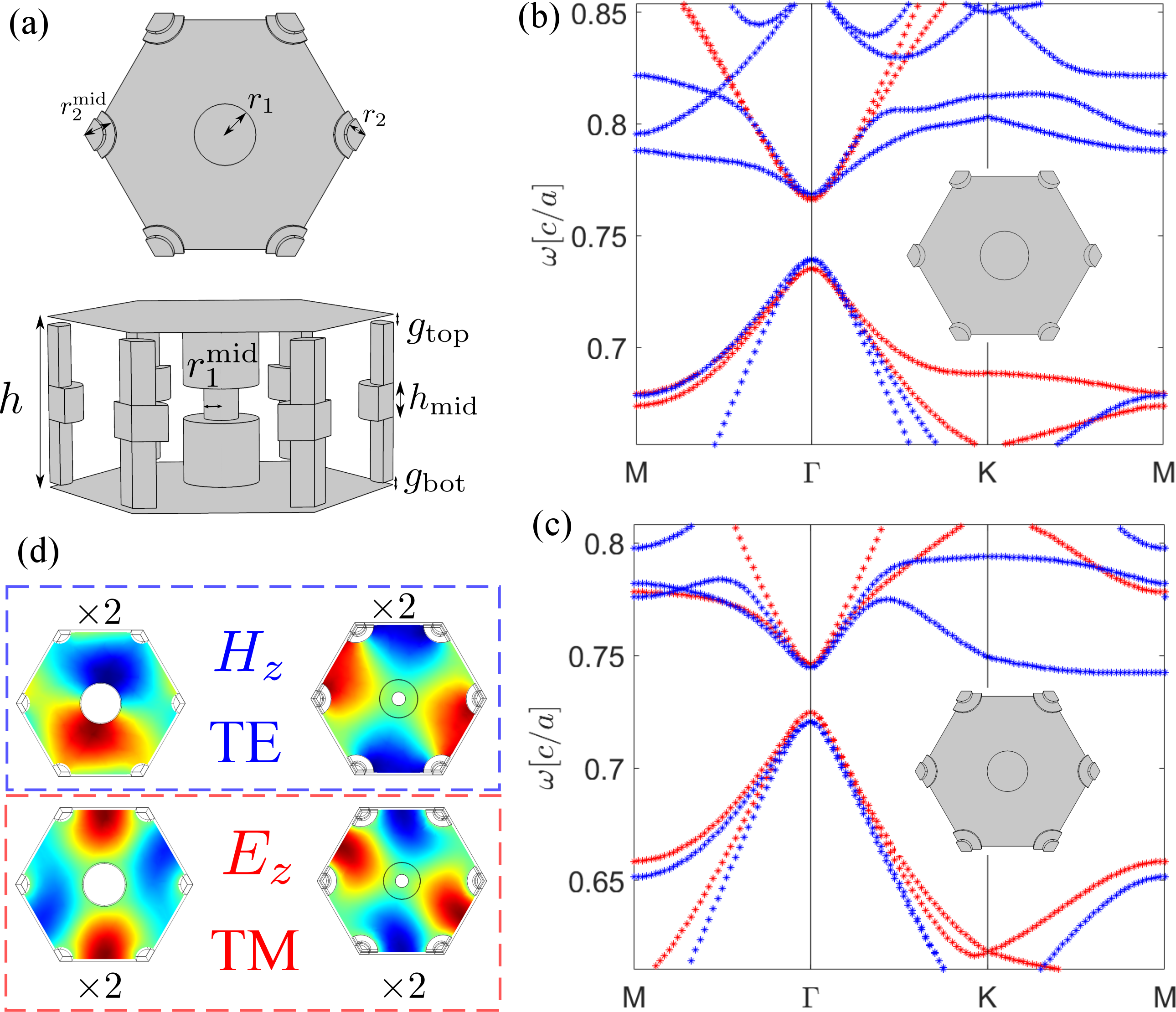}
    \caption{Emulation of spinful PHOTIs using a photonic crystal waveguide with Kekule distortion. \textbf{(a)} Top and side views of the PhC unit cell with inter-cell distance $a$. \textbf{(b,c)} Band structures for PHOTI (b) and spin-PHOTI (c) designs. TE (blue symbols) and TM (red symbols) are degenerate at the $\Gamma$ point. \textbf{(d)} Field profiles of the $\bf{p}$ and $\bf{d}$ orbitals bands below the band gap at the $\Gamma$ point for PHOTI (left column) and spin-PHOTI (right column) for TE (dashed blue box) and TM (dashed red box) electromagnetic modes. Parameters: symmetric air-gaps $g_{\rm top}=g_{\rm bot}$, see Table S2 for details.}\label{fig:Fig1}
\end{figure}

Next, we present a photonic platform that can be used for emulating a spinfull HOTI described by the   Hamiltonian given by Eq.~(\ref{eq:TB}). The structure shown in Fig.\ref{fig:Fig1}(a) is comprised of a photonic crystal (PhC) waveguide made of perfect electric conductors (PEC) elements sandwiched between two PEC plates~\cite{ma2015guiding,ma2017scattering,gladstone2019photonic}: see the SOM for the geometric parameters of all PhC designs. The PhC possesses $C_6$ symmetry, thereby allowing a Kekulé distortion~\cite{hou2007electron}, which opens a band gap at the $\Gamma$ point of the Brillouin Zone. In the case of symmetric air-gaps ($g_{\rm top} = g_{\rm bot}$) between the PhC elements and the plates, the mid-plane mirror symmetry $z\rightarrow -z$ ensures that the electromagnetic modes can be classified as either TE-like or TM-like, and that the two are decoupled from each other~\cite{joannopoulos2008molding}. The TE(TM) modes are distinguished by having a non-vanishing $H_z$ ($E_z$) field component at the $z=h/2$ midplane, as shown in Fig.\ref{fig:Fig1}(d).

In what follows, the TE(TM) nature of the modes will be used to emulate the $\uparrow$ ($\downarrow$) isospin components, while the mid-plane symmetry breaking $(\Delta g \equiv |g_{\text{top}} - g_{\text{bot}}| \neq 0)$ will be used to emulate the SOC~\cite{ma2015guiding,ma2017scattering}. While it is not possible to assign hopping amplitudes to the modes, we use the frequency ordering of the $\bf{p}=\left( p_x, p_y \right)$ and $\bf{d} = \left( d_{xy},d_{x^2-y^2} \right)$ orbitals with respect to the photonic bandgap at the $\Gamma$ point to classify the structures as trivial or topological~\cite{wu2015scheme,wu2016topological,jung2020nanopolaritonic}.

Using COMSOL simulations, we designed the unit cells shown in Fig.~\ref{fig:Fig1}(a) for two types of PhCs satisfying the following criteria for the TE(TM) modes: (i) frequency-degeneracy at the $\Gamma$-point for both PhCs, (ii) inverted photonic band structures (topological for one mode, trivial for the other) for a PHOTI-type PhC shown in Fig.~\ref{fig:Fig1}(b), and (iii) similar photonic band structures (topological or trivial for both modes) for the other type of a PhC. A PHOTI corresponding to (ii) -- trivial TE mode and a topological TM mode -- is shown in Fig.~\ref{fig:Fig1}(b), while a spin-PHOTI corresponding to (iii) -- topological for both modes -- is shown in Fig.~\ref{fig:Fig1}(c). Both stuctures have symmetric air-gaps, and the mid-plane orbital profiles of the TE ($H_z$) and TM ($E_z$) modes corresponding to the four bands below the bandgap are shown in Fig.~\ref{fig:Fig1}(d), where the left (right) column corresponds to a PHOTI (spin-PHOTI) structure.

\begin{figure}
    \centering
    \includegraphics[width=\columnwidth]{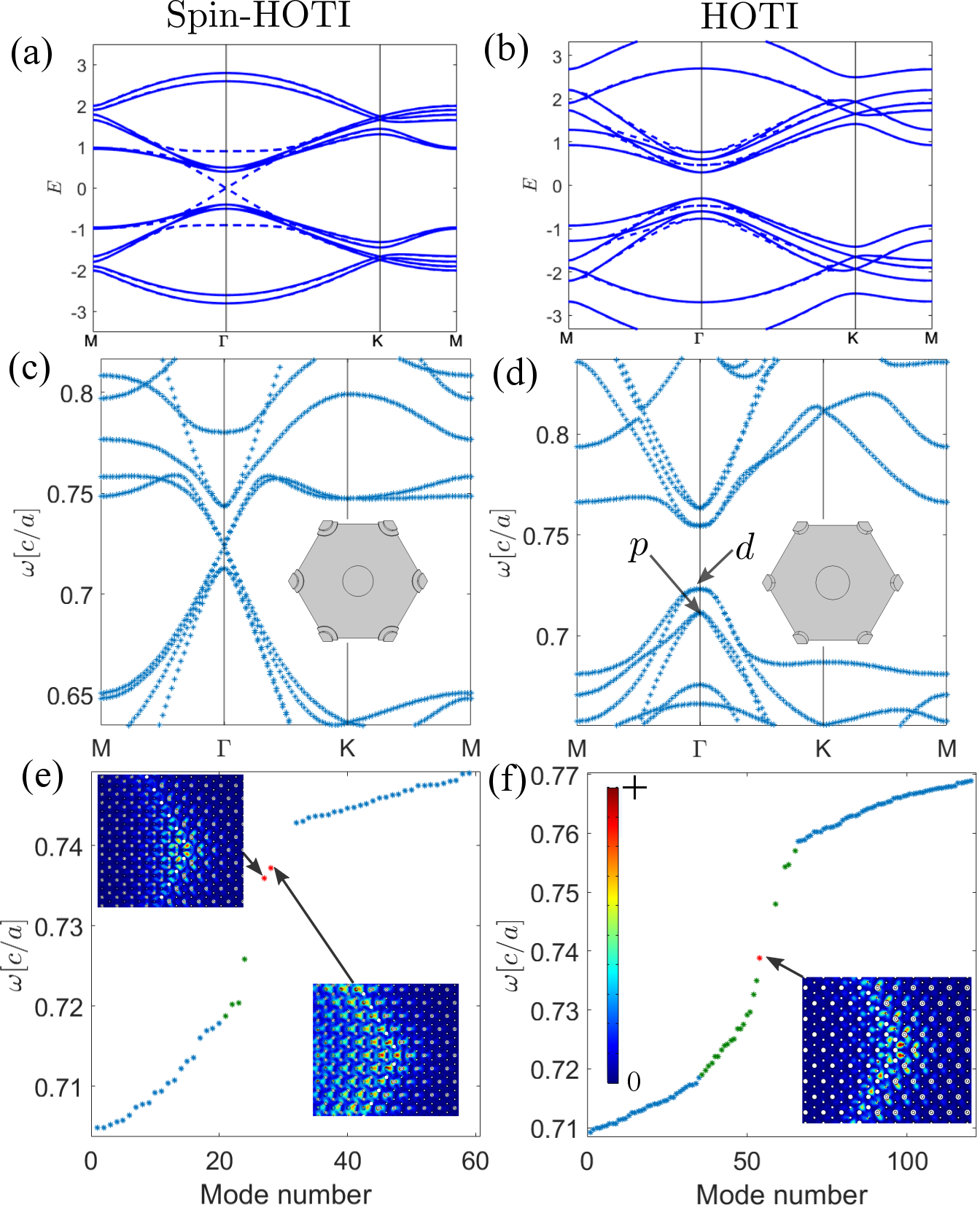}
    \caption{Comparison between HOTIs based on the TB model and their photonic counterparts: bulk bands (a-d) and corner states (e-f). Left column: spin-HOTI (a) and spin-PHOTI (c) phases. Right column: HOTI (b) and PHOTI (d) phases.  \textbf{(a)} Spin-HOTI bands without (with) the SOC term: solid (dashed: $\lambda_{\text{SOC}} = \lambda_{\text{SOC}}^{\rm th}$) lines. \textbf{(b)} Same as (a), but for HOTIs. \textbf{(c)} Photonic band structure for a spin-PHOTI phase with $\Delta g = \Delta g^{\rm (th)}$. Inset: top view of the unit cell. \textbf{(d)} Same as (c), but for a PHOTI phase. $p,d$ orbitals are marked with arrows. \textbf{(e)} Photonic corner energy spectrum of a spin-PHOTI and \textbf{(f)} HOTI interfacing a trivial phase at $120^{\circ}$ corner. Insets:  $|\mathbf{E}|^2$ (top of \textbf{(e)} and \textbf{(f)}) and $|\mathbf{H}|^2$ (bottom of \textbf{(e)}) profiles of the corner modes. Parameters: Table S2.}\label{fig:Fig4}
\end{figure}

To demonstrate that the designed PHOTI and spin-PHOTI structures emulate the corresponding phases described by the TB model given by Eq.(\ref{eq:TB}), we investigated phase transitions produced by the SOC-emulating air-gap asymmetry $\Delta g \neq 0$ by comparing the band structures calculated from the TB model shown in Figs.~\ref{fig:Fig4}(a),(b) to the first-principles photonic band structures shown in Figs.~\ref{fig:Fig4}(c),(d). No photonic bandgap closing occurs in the case of the PHOTI (Fig.~\ref{fig:Fig4}(d)) even as $\Delta g$ is progressively increased. On the other hand, complete bandgap closing for $\Delta g = \Delta g^{\rm (th)}$ is observed for the spin-PHOTI system, as indicated in Fig.~\ref{fig:Fig4}(c). Further increase in the effective SOC term ($\Delta g > \Delta g^{\rm (th)}$) reopens the bandgap~\cite{asboth2016short} and induces a phase transition from the spin-PHOTI to PHOTI phase (see SOM).

Next, we demonstrate that interfacing a photonic structure possessing quantized bulk invariants (either a PHOTI or spin-PHOTI) with a trivial PhC produces corner states with correct multiplicities predicted by the TB model of spinful HOTIs. For the PHOTI shown in Fig.~\ref{fig:Fig1}(c), we indeed find a single corner state marked by an arrow in Fig.\ref{fig:Fig4}(f) and plotted at a $120^{\circ}$ corner in the inset. We used edge roughening~\cite{jung2020nanopolaritonic} at the interface between corner-adjacent domain walls to increase the band gap between the edge modes, thereby further localizing the corner state.

Photonic analog of the degenerate ZCS pairs predicted by the TB model is constructed by interfacing a trivial PhC with a spin-PHOTI that was chosen to have the same parameters as the structure shown in Fig.~\ref{fig:Fig1}(c), but a smaller air-gap asymmetry $\Delta g < \Delta g^{\text{(th)}}$. Indeed, two nearly-degenerate corner states marked with arrows in Fig.\ref{fig:Fig4}(e) were found. Field intensities $|\mathbf{E}|^2$ and $|\mathbf{H}|^2$ of the two corner states  plotted in the inset show distinct spatial profiles and polarizations, which is equivalent to having spin texture. 

Several differences between the predictions of the simplified TB model and the results of the continuum electromagnetic calculation are notable. The edge modes shown in Fig.\ref{fig:Fig4}(e) and (f) as green dots appear only in the lower half of the bulk photonic bandgap (compare with Fig.~\ref{fig:TB_corner}(a),(c)) due to slight mismatch of the band gaps between the trivial and PHOTI phases. This mismatch is also responsible for the spectral shift of the corner state away from the mid-gap frequency and the lack of exact degeneracy between the corner states.

In summary, we have investigated higher-order topological insulating phases on a hexagonal lattice with spin-dependent Kekulé textures. When spin-flipping perturbations, such as spin-orbit coupling, are included in a tight-binding model, two types of insulating topological phases are predicted: a spin-HOTI possessing two independent fractionally-quantized bulk invariants $Q_c^{\uparrow \downarrow}$, and a HOTI possessing just one such invariant $Q_c = Q_c^{\uparrow} + Q_c^{\downarrow}$. A bulk-boundary correspondence between such bulk invariants and the existence of corner states is established, and their photonic analogues are proposed. Our results present an opportunity for future development of novel photonic devices with active switching of their topological corner states by controlling the mid-plane mirror symmetry, thereby inducing a topological phase transition. Condensed matter realization of the spin-HOTI phase with $\Delta_\uparrow = \Delta_\downarrow$ is possible using a platform such as CO molecules deposited on Cu~\cite{freeney2020edge}. An effective SOC can be introduced by applying a magnetic field, which would introduce a phase transition to a quantum Hall phase instead of a QSH phase.

This work was supported by the Office of Naval Research Award No. N00014-21-1-2056, National Science Foundation Award No. NNCI-1542081, and the Army Research Office Award W911NF2110180. M. J. also acknowledges the support from
the Kwanjeong Fellowship from the Kwanjeong Educational
Foundation.

\FloatBarrier

\bibliography{TopologicalSpinCornerStates_GS}

\pagebreak
\widetext
\begin{center}
\textbf{\large Supplemental Materials: Title for main text}
\end{center}
\setcounter{equation}{0}
\setcounter{figure}{0}
\setcounter{table}{0}
\setcounter{page}{1}
\makeatletter
\renewcommand{\theequation}{S\arabic{equation}}
\renewcommand{\thefigure}{S\arabic{figure}}
\renewcommand{\bibnumfmt}[1]{[S#1]}

\section{High symmetry points mode profiles and calculation of the spin polarized corner charge}
In the main text we presented mode profiles only at $\Gamma$ because that is where the topological band inversion occurs. However, it is important to inspect all the relevant mode profiles of the TE-like and TM-like modes at both the $\Gamma$ and $M$ points to verify that our PhC designs indeed have the expected quantized corner charge. We will do that here for the PHOTI phase of Fig.~3(b) and count the number of modes below the band gap with $+1$ eigenvalue to $C_2$ rotation. The same procedure can be repeated for all the other designs. 

\begin{figure}
    \centering
    \includegraphics[width=120mm]{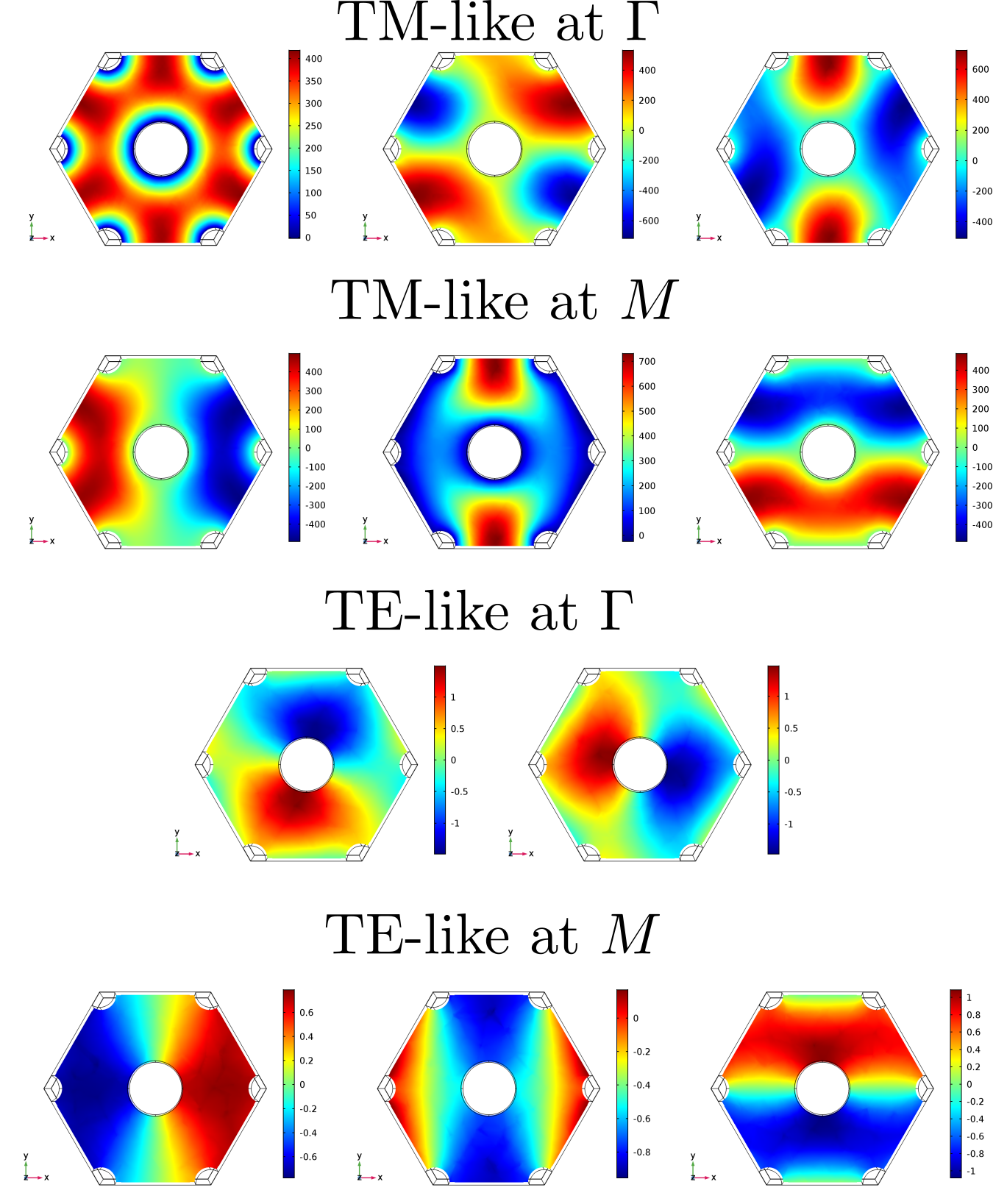}
    \caption{Mode profiles for TM-like and TE-like modes of relevant bands below the band gap, at $\Gamma$ and $M$ points. $H_z$~($E_z$) field at $z=h/2$ plotted for TE-like (TM-like) modes.}
    \label{fig:modeProfiles}
\end{figure}

We present the relevant field profiles at Fig.~\ref{fig:modeProfiles}. We only present modes that fit our definition of TE-like and TM-like from the main text. Notably, there is a missing TE-like mode at $\Gamma$. The mode does not actually disappear, but rather its profile no longer adheres to a TE-like mode profile due to its low frequency. If the mode retained its profile it could only have an eigenvalue of +1 to $C_2$, being the ground mode of that polarization.

For the TM-like modes we have 3 modes with +1 eigenvalues to $C_2$ at $\Gamma$ and only 1 at $M$. Using the mode profiles to calculate $[M]$ defned by Benalcazar et al.~\cite{benalcazar2019quantization} we find that $[M]_\uparrow = -2$ and $Q_c^{\uparrow} = [M]/4 = 1/2 \text{ (mod 1)}$. Similarly, assuming the missing TE-like mode has a symmetric profile we have 0 modes with eigenvalues +1 to $C_2$ at $\Gamma$ and 1 at $M$. Therefore for the TE-like modes $[M]_\downarrow = 0$ and $Q_c^{\downarrow} = 0 \text{ (mod 1)}$. This is the expected result for this realization of the HOTI phase with no SOC.

\section{Momentum space tight binding Hamiltonian}
Fourier transforming the tight binding model of Eq.~(1) from the main text, one finds that the Hamiltonian can be written in four blocks. The two diagonal blocks, one for each spin, are the standard Kekulé texture Hamiltonian~\cite{jung2020nanopolaritonic} and the off diagonal blocks describe the SOC:
\begin{equation}
    H = 
    \begin{pmatrix}
    H_\uparrow & H_{\text{SOC}} \\
    H^\dagger_{\text{SOC}} & H_\downarrow
    \end{pmatrix},
    \label{eq:Hamiltonian}
\end{equation}

where 

\begin{equation}
    H_{\uparrow,\downarrow} = 
    \begin{pmatrix}
    0 & -t_{\text{in}}^{\uparrow,\downarrow} & 0 & -t_{\text{out}}^{\uparrow,\downarrow}e^{-i\mathbf{k}\cdot\mathbf{a}_1} & 0 & -t_{\text{in}}^{\uparrow,\downarrow} \\
    -t_{\text{in}}^{\uparrow,\downarrow} & 0 & -t_{\text{in}}^{\uparrow,\downarrow} & 0 & -t_{\text{out}}^{\uparrow,\downarrow}e^{i\mathbf{k}\cdot\mathbf{a}_2} & 0\\
    0 & -t_{\text{in}}^{\uparrow,\downarrow} & 0 & -t_{\text{in}}^{\uparrow,\downarrow} & 0 & -t_{\text{out}}^{\uparrow,\downarrow}e^{-i\mathbf{k}\cdot\mathbf{a}_3} \\
    -t_{\text{out}}^{\uparrow,\downarrow}e^{i\mathbf{k}\cdot\mathbf{a}_1} & 0 & -t_{\text{in}}^{\uparrow,\downarrow} & 0 & -t_{\text{in}}^{\uparrow,\downarrow} & 0 \\
    0 & -t_{\text{out}}^{\uparrow,\downarrow}e^{-i\mathbf{k}\cdot\mathbf{a}_2} & 0 & -t_{\text{in}}^{\uparrow,\downarrow} & 0 & -t_{\text{in}}^{\uparrow,\downarrow} \\
    -t_{\text{in}}^{\uparrow,\downarrow} & 0 & -t_{\text{out}}^{\uparrow,\downarrow}e^{i\mathbf{k}\cdot\mathbf{a}_3} & 0 & -t_{\text{in}}^{\uparrow,\downarrow} & 0
    \end{pmatrix},
    \label{eq:Hupdown}
\end{equation}

\begin{equation}
    \tiny
    H_{\text{SOC}} = i\frac{\lambda_\text{SOC}}{3\sqrt{3}}
    \begin{pmatrix}
    0 & 0 & -1-e^{-i\mathbf{k}\cdot \mathbf{a}_1}-e^{i\mathbf{k}\cdot \mathbf{a}_3} & 0 & 1+e^{-i\mathbf{k}\cdot \mathbf{a}_1}+e^{i\mathbf{k}\cdot \mathbf{a}_2} & 0 \\
    0 & 0 & 0 & -1-e^{-i\mathbf{k}\cdot \mathbf{a}_1}-e^{i\mathbf{k}\cdot \mathbf{a}_2} & 0 & 1+e^{-i\mathbf{k}\cdot \mathbf{a}_3}+e^{i\mathbf{k}\cdot \mathbf{a}_2} \\
    1+e^{-i\mathbf{k}\cdot \mathbf{a}_3}+e^{i\mathbf{k}\cdot \mathbf{a}_1} & 0 & 0 & 0 & -1-e^{-i\mathbf{k}\cdot \mathbf{a}_3}-e^{i\mathbf{k}\cdot \mathbf{a}_2} & 0\\
    0 & 1+e^{-i\mathbf{k}\cdot \mathbf{a}_2}+e^{i\mathbf{k}\cdot \mathbf{a}_1} & 0 & 0 & 0 & -1-e^{-i\mathbf{k}\cdot \mathbf{a}_1}-e^{i\mathbf{k}\cdot \mathbf{a}_3} \\
    -1-e^{-i\mathbf{k}\cdot \mathbf{a}_2}-e^{i\mathbf{k}\cdot \mathbf{a}_1} & 0 & 1+e^{-i\mathbf{k}\cdot \mathbf{a}_2}+e^{i\mathbf{k}\cdot \mathbf{a}_3} & 0 & 0 & 0 \\
    0 & -1-e^{-i\mathbf{k}\cdot \mathbf{a}_2}-e^{i\mathbf{k}\cdot \mathbf{a}_3} & 0 & 1+e^{-i\mathbf{k}\cdot \mathbf{a}_3}+e^{i\mathbf{k}\cdot \mathbf{a}_1} & 0
    \end{pmatrix},
    \normalsize
    \label{eq:HSOC}
\end{equation}

$\mathbf{a}_1 = a(1,0)$, $\mathbf{a}_2 = a(-1/2,\sqrt{3}/2)$ and $\mathbf{a}_3 = (-1/2,-\sqrt{3}/2)$. Defining a $C_2$ operator~\cite{jung2020nanopolaritonic} for a single spin as 
\begin{equation}
    C_2 =
    \begin{pmatrix}
    0 & 0 & 0 & 1 & 0 & 0 \\
    0 & 0 & 0 & 0 & 1 & 0 \\
    0 & 0 & 0 & 0 & 0 & 1 \\
    1 & 0 & 0 & 0 & 0 & 0 \\
    0 & 1 & 0 & 0 & 0 & 0 \\
    0 & 0 & 1 & 0 & 0 & 0 \\
    \end{pmatrix},
\end{equation}
one can verify that the operator $C_2 \otimes I_{2\times 2}$ is a symmetry operator for Eq.~(\ref{eq:Hamiltonian}) at both $\Gamma$ and $M$, which means they remain rotation invariant momenta points. The same procedure can also be applied to the $C_3$ operator:
\begin{equation}
    C_3 =
    \begin{pmatrix}
    0 & 0 & 1 & 0 & 0 & 0 \\
    0 & 0 & 0 & 1 & 0 & 0 \\
    0 & 0 & 0 & 0 & 1 & 0 \\
    0 & 0 & 0 & 0 & 0 & 1 \\
    1 & 0 & 0 & 0 & 0 & 0 \\
    0 & 1 & 0 & 0 & 0 & 0 \\
    \end{pmatrix},
\end{equation}
and $C_3\otimes I_{2\times 2}$, which shows that $\Gamma$ and $K$ remain rotation invariant momenta. This is why the addition of SOC as a perturbation does not change the eigenvalues of the system and why the addition of SOC does not change the total corner charge $Q_c$.

When setting $t_{\text{in}}^{\uparrow} = t_{\text{in}}^{\downarrow}$ and $t_{\text{out}}^{\uparrow} = t_{\text{out}}^{\downarrow}$, meaning the spin subspaces are degenerate, transforming this Hamiltonian according to the unitary transformation $H\rightarrow \exp{(i\pi s_y / 4)} H \exp({-i \pi s_y /4)}$, where $\hat{s}_y$ is the second Pauli matrix, results in a block diagonal Hamiltonian, meaning we can still define a spin polarized corner charge for each subspace using Eq.~(2) of the main text without resorting to Eq.~(3).

\section{Derivation of phase transition condition $\lambda^2_\text{SOC} = \Delta_\uparrow\Delta_\downarrow$ and coupling of bands through SOC}
Without SOC the Hamiltonian of Eq.~(\ref{eq:Hamiltonian}) has two independent spin subspaces. Without loss of generality we assume the topological phase. Additionally, we operate at the $\Gamma$ point because that is where the band gap closes. The spin up subspace will have two $p$ orbital modes at $\Delta_\uparrow$ and two $d$ orbital modes at $-\Delta_\uparrow$. Likewise, the spin down subspace will have two $p$ orbital modes at $\Delta_\downarrow$ and two $d$ orbitals at $-\Delta_\downarrow$. The SOC perturbation couples spin up $p$ $(d)$ orbitals to spin down $p$ $(d)$ orbitals but does not couple orbitals of different types, because it respects $C_6$ symmetry, and the coupling strength is $\lambda_\text{SOC}$. This can be seen by transforming $H_\text{SOC}$ to the orbital basis~\cite{jung2020nanopolaritonic} defined according to $\left[ s, p_x+ip_y, p_x-ip_y, d_{xy}+id_{x^2-y^2}, d_{xy}-id_{x^2-y^2}, f \right]$, where:
\begin{equation}
\begin{split}
    s &= [1,1,1,1,1,1]/\sqrt{6}, \\
     p_x &=[1,0,0,-1,0,0]/\sqrt{2}, \\
     p_y &=[0,1,1,0,-1,-1]/2,\\
     d_{xy} &=[2,-1,-1,2,-1,-1]/\sqrt{12},\\
     d_{x^2-y^2}&=[1,-1,-1,1,-1,-1]/\sqrt{6},\\
     f&=[1,-1,1,-1,1,-1]/\sqrt{6}.
     \end{split}
\end{equation}

Performing this transformation one finds $H_{\text{SOC}}=\lambda_{\text{SOC}}B^{\dagger}\text{diag}(0,-1,-1,1,1,0) B$, where $B=[s, p_x+ip_y, p_x-ip_y, d_{xy}+id_{x^2-y^2}, d_{xy}-id_{x^2-y^2}, f]$.
Therefore the eigenenergies after the SOC perturbation are the eigenvalues of the matrices 
\begin{equation}
    \begin{pmatrix}
    \Delta_\uparrow & \lambda_\text{SOC} \\
    \lambda_\text{SOC} & \Delta_\downarrow,
    \end{pmatrix},
    \begin{pmatrix}
    -\Delta_\uparrow & \lambda_\text{SOC} \\
    \lambda_\text{SOC} & -\Delta_\downarrow,
    \end{pmatrix}
\end{equation}
for the qudrupole and dipole modes respectively. Solving for the eigenvalues of these matrices we find that the quadrupole energies are 
\begin{equation}
    E_{d\pm} = -\frac{\Delta_\uparrow + \Delta_\downarrow}{2}\pm \sqrt{\lambda_\text{SOC}^2 + \left(\frac{\Delta_\uparrow - \Delta_\downarrow}{2}\right)^2}
\end{equation}
and the dipole energies are 
\begin{equation}
    E_{p\pm} = \frac{\Delta_\uparrow + \Delta_\downarrow}{2}\pm \sqrt{\lambda_\text{SOC}^2 + \left(\frac{\Delta_\uparrow - \Delta_\downarrow}{2}\right)^2}.
\end{equation}
Setting $ E_{d-} = E_{p+} = 0$ gives us the band gap closing condition of $\lambda_\text{SOC}^2 = \Delta_\uparrow \Delta_\downarrow$.

\section{Derivation of photonic effective  Hamiltonian and chiral symmetry}
In this section we show the correspondence between the photonic PEC bianisotropic waveguide and the tight binding model of Eqs.~(\ref{eq:Hamiltonian}-\ref{eq:HSOC}) and discuss its limitations. We first deal with the TE-like and TM-like subspaces independently. We designed our waveguides such that they all posses eigenmodes with field profiles that have $s$, $p_x$, $p_y$,$d_{xy}$,$d_{x^2-y^2}$ and $f$ orbital profiles and are separated well enough in energy from any additional bands that may interact with the relevant bands. Imposing $C_6$ symmetry on each subspace, i.e. $C_6 H C_6 = H$ we find that $H$ is a circulant matrix at the rotation invariant momenta of $\Gamma$ and $M$. This holds exactly for the TB model of Eq.~(\ref{eq:Hupdown}) as well as the photonic waveguide. 

Chiral symmetry of the form $S H S = -H$ with $S=\text{diag}(1,-1,1,-1,1,-1)$ applies to $H_{\uparrow,\downarrow}$ but for the photonic system it is not exact if one considers possible hopping terms that are not exclusively nearest neighbor. We therefore have to assume that if we take the photonic eigenmodes and transform them to the sublattice basis, where each mode is localized at an effective lattice site in the unit cell~\cite{jung2020nanopolaritonic}, the overlap integrals between next-nearest neighbors modes of the same polarization are negligible compared to nearest neighbor overlap integrals. Phenomenologically, this seems to apply because the frequency of the corner modes is very close to the mid-gap frequency in all our designs. We note that one can design photonic systems where chiral symmetry breaks (see supplementary information of~\cite{jung2020nanopolaritonic}), hence chiral symmetry for the photonic system is by design, not an intrinsic symmetry like in the lattice models. 

The meaning of chiral symmetry in the lattice model is that the lattice can be divided into two sublattices, one containing the odd index sites and the other the even index sites. In the orbital basis the meaning of the chiral symmetry operator is seen by its action on the basis vectors $\left| s\right>$, $\left| p_\pm\right>$, $\left| d_\pm\right>$, $\left| f\right>$:
\begin{equation}
\begin{split}
    S\left|s\right> = \left|f\right>,\\
    S\left|f\right> = \left|s\right>,\\
    S\left|p_\pm\right> = \left|d_\pm\right>,\\
    S\left|d_\pm\right> = \left|p_\pm\right>,
\end{split}
\end{equation}
where $\left|p_\pm\right> = \left|p_x\right> \pm i \left|p_y\right>$ and $\left|d_\pm\right> = \left|d_{xy}\right> \pm i \left|d_{x^2-y^2}\right>$.

Assuming chiral symmetry applies to the photonic waveguide, one finds that the even indices of the circulant matrix $H_{\uparrow,\downarrow}$ at $\Gamma$ and $M$ must all be 0. The form of the effective Hamiltonian is therefore:
\begin{equation}
H_{\uparrow,\downarrow} =
    \begin{pmatrix}
    0 & c_1 & 0 & c_3 & 0 & c_5 \\
    c_5 & 0 & c_1 & 0 & c_3 & 0 \\
    0 & c_5 & 0 & c_1 & 0 & c_3 \\
    c_3 & 0 & c_5 & 0 & c_1 & 0 \\
    0 & c_3 & 0 & c_5 & 0 & c_1 \\
    c_1 & 0 & c_3 & 0 & c_5 & 0 \\
    \end{pmatrix}.
\end{equation}
Demanding bosonic time-reversal symmetry $KHK=H$ where $K$ is the complex conjugation operator means all elements of the Hamiltonian are real at the time-reversal invariant momenta $\Gamma$ and $M$. Demanding that the system is lossless constrains the Hamiltonian to be Hermitian, i.e. $H^\dagger = H$, and since it is also real $H^T=H$, hence $c_1 = c_5$, giving the final form of \begin{equation}
H_{\uparrow,\downarrow} =
    \begin{pmatrix}
    0 & c_1 & 0 & c_3 & 0 & c_1 \\
    c_1 & 0 & c_1 & 0 & c_3 & 0 \\
    0 & c_1 & 0 & c_1 & 0 & c_3 \\
    c_3 & 0 & c_1 & 0 & c_1 & 0 \\
    0 & c_3 & 0 & c_1 & 0 & c_1 \\
    c_1 & 0 & c_3 & 0 & c_1 & 0 \\
    \end{pmatrix}.
\end{equation}
which is identical to the values obtained by Eq.~(\ref{eq:Hupdown}) at $\Gamma$ and $M$ if one sets $c_1 = -t_\text{in}$ and $c_3 = -t_\text{out}$.

Next, we turn to consider the effects of breaking mid-plane mirror symmetry which couples the TE-like modes and the TM-like modes. From Slater cavity perturbation theory~\cite{ma2015guiding}, we find the effect of the perturbation on the $H_{\uparrow,\downarrow}$ diagonal blocks is simply a frequency shift of the TE-like and TM-like subspaces that can be engineered to be the same magnitude for both polarizations by optimizing the energy overlap of the TE-like modes and TM-like modes after the introduction of the mirror symmetry breaking perturbation. The perturbation also introduces off-diagonal blocks, whose form we analyze below.

Imposing $C_6$ symmetry of the form $\Tilde{C_6} = C_6 \otimes I_{2\times2}$, i.e. $\Tilde{C6} H \Tilde{C_6} = H$, we find that the off-diagonal blocks $H_\text{SOC}$ are circulant matrices, just like the diagonal blocks. Since the systems is still lossless, the combined Hamiltonian is still Hermitian, so the off-diagonal blocks are Hermitian conjugates of each other, and the system still has bosonic time-reversal symmetry, so its elements are all real. We can therefore construct one off-diagonal block and the other is determined according to the first. 

Assuming the frequency shift of both polarizations is 0, then due to time-reversal property of a bianistropic waveguide with real $\underline{\underline{\varepsilon}}$,$\underline{\underline{\mu}}$ and $\underline{\underline{\chi}}$~\cite{xiong2017classification} we know that a chirality operator $\Tilde{S}$ must exist such that $\Tilde{S} H \Tilde{S} = -H$. Furthermore, we know this operator must flip the sign of the TM-like modes because it flips the sign of the transverse magnetic field~\cite{xiong2017classification}. $\Tilde{S}$ must therefore take the form $\Tilde{S} = S\otimes\sigma_z$, where $S$ is some operator acting on the TE-like and TM-like subspaces separately. Using the assumption of chiral symmetry for $H_{\uparrow,\downarrow}$ before the introduction of mirror-symmetry breaking we find that $S$ can only be identical to $S=\text{diag}(1,-1,1,-1,1,-1)$ introduced before.

The geometric meaning of this chiral symmetry in the sublattice basis is that the lattice can still be divided into two sublattices even with the introduction of spin and hopping terms that involve spin. The first new sublattice is the spin up / TE-like modes at sites 1,3,5 and spin down / TM-like modes at sites 1,3,5 and the second sublattice is spin down / TM-like at sites 2,4,6 and TE-like at sites 2,4,6.

Imposing $\Tilde{S} H \Tilde{S} = -H$ constraints the off-diagonal blocks to obey $S H_\text{SOC} S = H_\text{SOC}$, where the negative sign is gone due to the $\sigma_z$ component of $\Tilde{S}$ for the off-diagonal blocks. One then finds that the circulant matrix $H_\text{SOC}$ odd index elements vanish. After a change of basis $S_x\rightarrow -S_y$ all elements become imaginary. Therefore at $\Gamma$ and $M$ it has the form:
\begin{equation}
H_{\text{SOC}} = i
    \begin{pmatrix}
    c_0 & 0 & c_2 & 0 & c_4 & 0 \\
    0 & c_0 & 0 & c_2 & 0 & c_4 \\
    c_4 & 0 & c_0 & 0 & c_2 & 0 \\
    0 & c_4 & 0 & c_0 & 0 & c_2 \\
    c_2 & 0 & c_4 & 0 & c_0 & 0 \\
    0 & c_2 & 0 & c_4 & 0 & c_0 \\
    \end{pmatrix}.
\end{equation}
Kane-Mele SOC of Eq.(\ref{eq:HSOC}) is a particular example of such allowed coupling matrix at $\Gamma$ and $M$, which is why the photonic model and the TB model are so similar in their behavior.

\section{Additional phase diagrams}
\begin{figure}
    \centering
    \includegraphics[width=180mm]{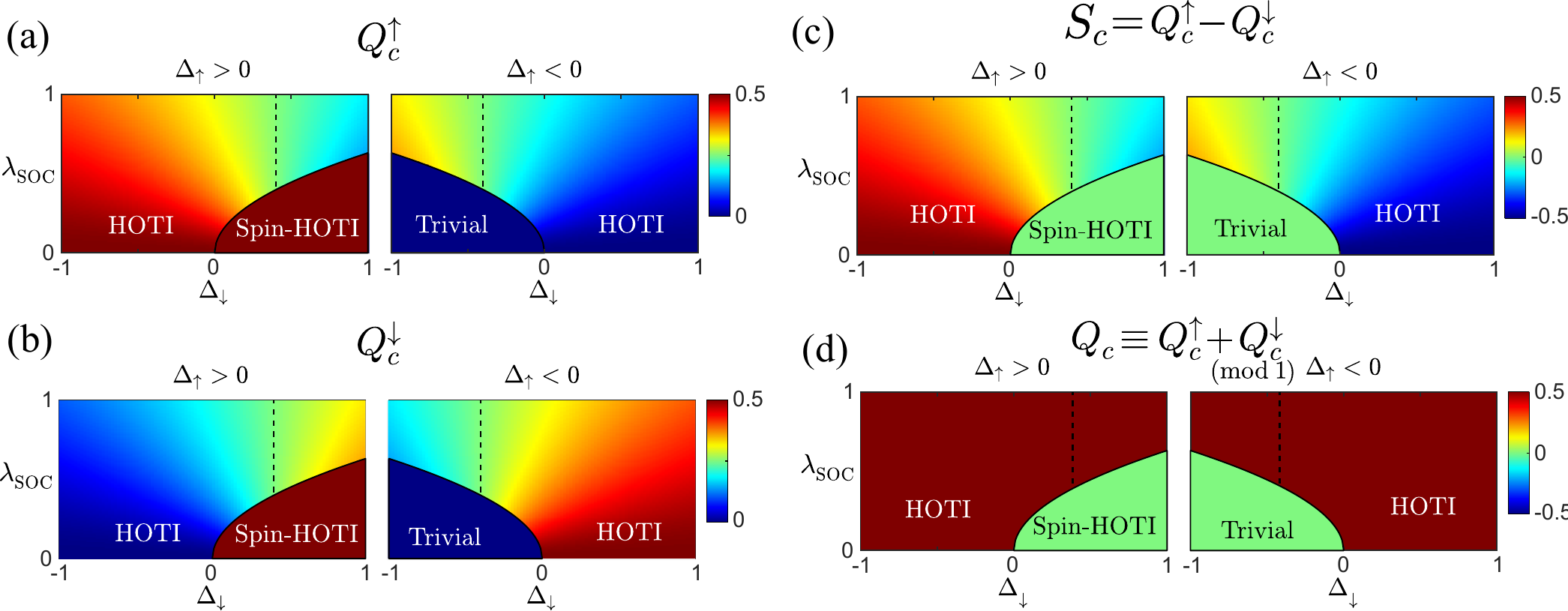}
    \caption{Phase diagrams with constant $\Delta_\uparrow$ with additional color content for bulk quantities not shown in the main text. \textbf{(a)}Color content is $Q_c^\uparrow$. Black dashed line marks the QSH phase. \textbf{(b)} Color is $Q_c^\downarrow$. \textbf{(c)} Color is $S_c$. \textbf{(d)} Color is $Q_c$. Note that the color bars for different sub-figures have different ranges.}
    \label{fig:S4}
\end{figure}

\begin{figure}
    \centering
    \includegraphics[width=120mm]{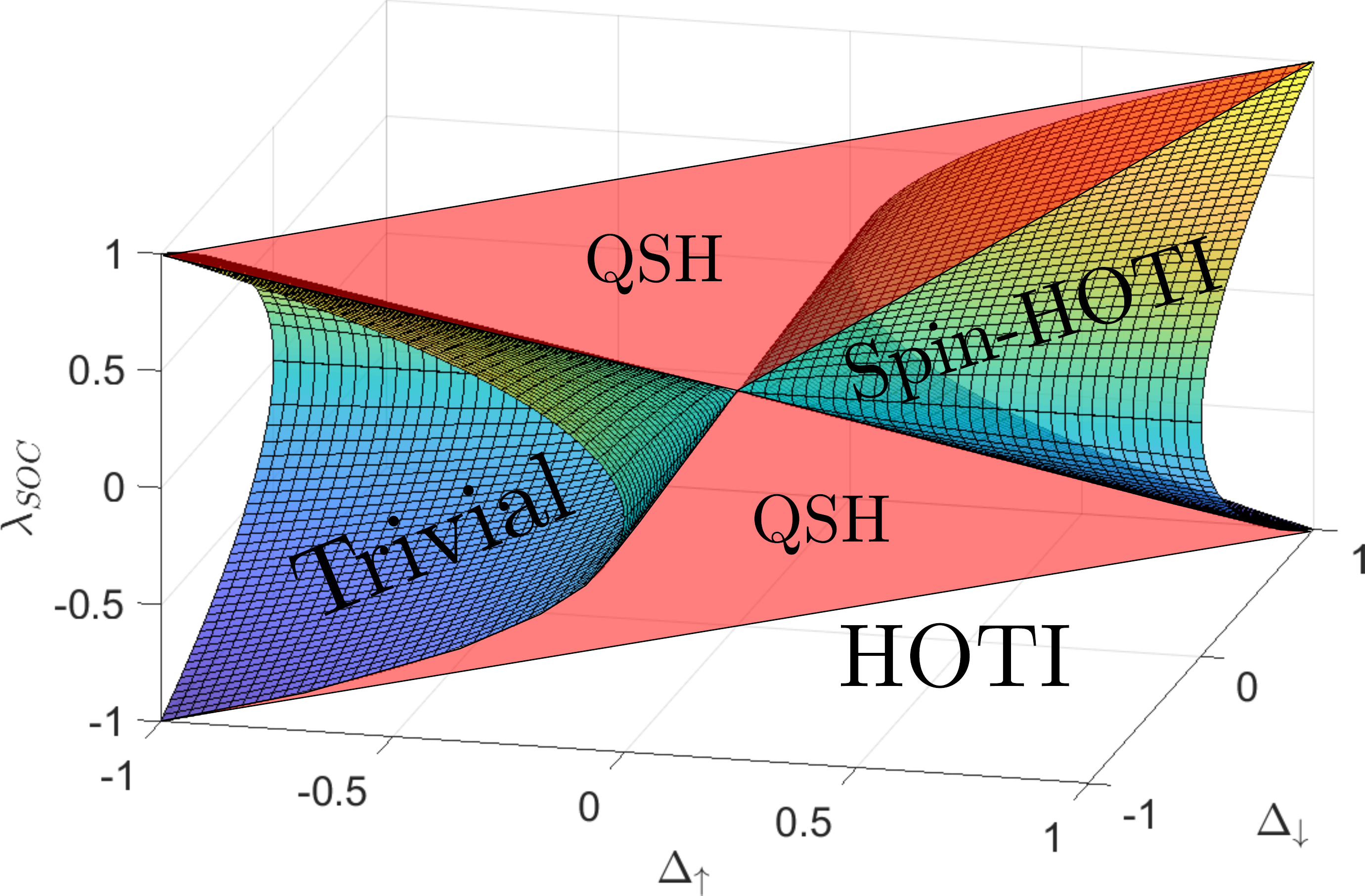}
    \caption{3D phase diagram. Phase boundaries are plotted using meshed surface. HOTI phase captures all the volume that is not contained in the trivial and spin-HOTI surfaces. QSH planes are in transparent red.}
    \label{fig:3dPhase}
\end{figure}

We chose to present $Q_c^\uparrow$ as the color for Fig.~1(c), but there is some additional value in presenting additional phase diagrams with the color encoding the additional bulk quantities we have defined in the main text. Fig.~\ref{fig:S4}(a) shows the phase diagram with $Q_c^\uparrow$ as the color. In Fig.~\ref{fig:S4}(b) we use $Q_c^\downarrow$ as the color content. We can see both are quantized at 1/2 for the topological phase and 0 for the trivial phase, but lose quantization for the HOTI phases as SOC is added. In Fig.~\ref{fig:S4}(c) we plot $S_c$ as the color content and notice that it retains $S_c = 0$ along the black dashed QSH line. This is expected since the spin subspaces are identical for that parameter choice and therefore $Q_c^\uparrow = Q_c^\downarrow$. Comparing to Fig.~\ref{fig:S4}(a) one can see that $S_c$ is a function of only one of $Q_c^\uparrow$ or $Q_c^\downarrow$, i.e. they are not independent. This is because for the HOTI phase $Q_c^\downarrow = 1/2 - Q_c^\uparrow$ due to the quantization of $Q_c$. In Fig.~\ref{fig:S4}(d) we plot $Q_c$ as the color content, which is always quantized for all phases. Additionally, we plot the 3D phase diagram in Fig.~\ref{fig:3dPhase}, whose cut planes are plotted in Fig.~\ref{fig:S4}.

\section{Quantization of $Q_c^{\uparrow}$ and $Q_c^{\downarrow}$ in spin-HOTI phase with a finite $\lambda_{\text{SOC}}$}
In this section, we provide a brief proof on how the spin-projected corner charge indices $Q_c^{s_z=\uparrow}$ and $Q_c^{s_z=\downarrow}$ are well-quantized in the spin-HOTI phase $Q_c^{\uparrow\downarrow}=\frac{1}{2}$ and in the trivial phase $Q_c^{\uparrow\downarrow}=0$ even in the presence of an off-diagonal spin-orbit coupling ($H_{\text{SOC}}\otimes s_x$). In the section III, we established that $H_{\text{SOC}}=\left[\lambda_{\text{SOC}} B^{\dagger}\text{diag}(0,-1,+1,-1,+1,0)B \right]$, where $B=[s,p_-, p_+, d_-, d_+, f]$ ($p_{\pm} = p_x \pm p_y$, $d_{\pm} = d_{xy} \pm d_{x^2-y^2}$). In other words, $H_{\text{SOC}}\otimes s_x$ can induce coupling only between $\left|{p_{\pm}, \uparrow}\right>$ and $\left|{p_{\pm}, \downarrow}\right>$ or between $\left| {d_{\pm}, \uparrow}\right>$ and $\left|{d_{\pm} , \downarrow}\right>$.

In the spin-topological phase ($\Delta_{\uparrow\downarrow}>0$), before we introduce a finite $\lambda_{\text{SOC}}$, we have $\left| {s,\uparrow} \right>$, $\left|{d_\pm,\uparrow} \right>$, $\left|{s,\downarrow}\right>$, $\left|{d_\pm,\downarrow}\right>$ orbitals below the bandgap and $\left|{f,\uparrow}\right>$, $\left|{p_\pm,\uparrow}\right>$, $\left|{f,\downarrow}\right>$, $\left|{p_\pm,\downarrow}\right>$ orbitals above the bandgap at $\bf{\Gamma}$-point, thereby featuring $\#\Gamma_{\uparrow,\downarrow}=3$ (for both spin subspaces, there are three eigenstates with +1 eigenvalue to $C_2$ rotation below the band gap). We can see that a finite $H_{\text{SOC}}$ term cannot induce any coupling between the states that are placed across the bandgap. Then, given a finite $\lambda_{\text{SOC}}$, the eigenstates at $\bf{\Gamma}$-point below the bandgap will be $\left|{s,\uparrow}\right>$, $\left|{s,\downarrow}\right>$, $A\left|{d_+,\uparrow}\right>+B\left|{d_+,\downarrow}\right>$, $B\left|{d_+,\uparrow}\right>-A\left|{d_+,\downarrow}\right>$, $C\left|{d_-,\uparrow}\right>+D\left|{d_-,\downarrow}\right>$, and $D\left|{d_-,\uparrow}\right>-C\left|{d_-,\downarrow}\right>$. The linear coefficients $A$ and $B$ are given by solving the eigenbasis for $[\Delta_\uparrow, \lambda_{\text{SOC}};\lambda_{\text{SOC}}, \Delta_\downarrow]$, and $C$ and $D$ for $[\Delta_\uparrow, -\lambda_{\text{SOC}};-\lambda_{\text{SOC}}, \Delta_\downarrow]$. To be specific, $\frac{A}{B}=-\frac{C}{D}=\frac{\Delta_\uparrow-\Delta_\downarrow+\sqrt{(\Delta_\uparrow-\Delta_\downarrow)^2+4\lambda_{\text{SOC}}^2}}{2\lambda_{\text{SOC}}}$ with $A^2+B^2=1$ and $C^2+D^2=1$ (they can be set to be real numbers). Then, we get $\#\Gamma_{\uparrow\downarrow}=1+A^2+B^2+C^2+D^2=3$. Therefore, upon a finite $\lambda_{\text{SOC}}$, $\#\Gamma_{\uparrow\downarrow}$ is preserved to be at the same value obtained with vanishing $\lambda_{\text{SOC}}$. This same analysis applies for the trivial phase as well that $\#\Gamma_{\uparrow\downarrow}=1$ is preserved with or without $\lambda_{\text{SOC}}$.

Basically, since the Kane-mele type spin-orbit coupling induces coupling only between the same orbital types, in the trivial or spin-HOTI phase, the exchange of wavefunction amplitudes is happening only within the bands that are below the bandgap or only within the bands that are above the bandgap. Thus, the sum of spin-projected wavefunction amplitudes over all eigenstates (with $C_2$ rotation eigenvalue of +1) below the bandgap remains the same unless $\lambda_{\text{SOC}}$ is too strong to cause the band inversion.

\section{Fractional corner-localized charge and corner-localized spin as boundary observables}
\begin{figure}
    \includegraphics[width=0.9\columnwidth]{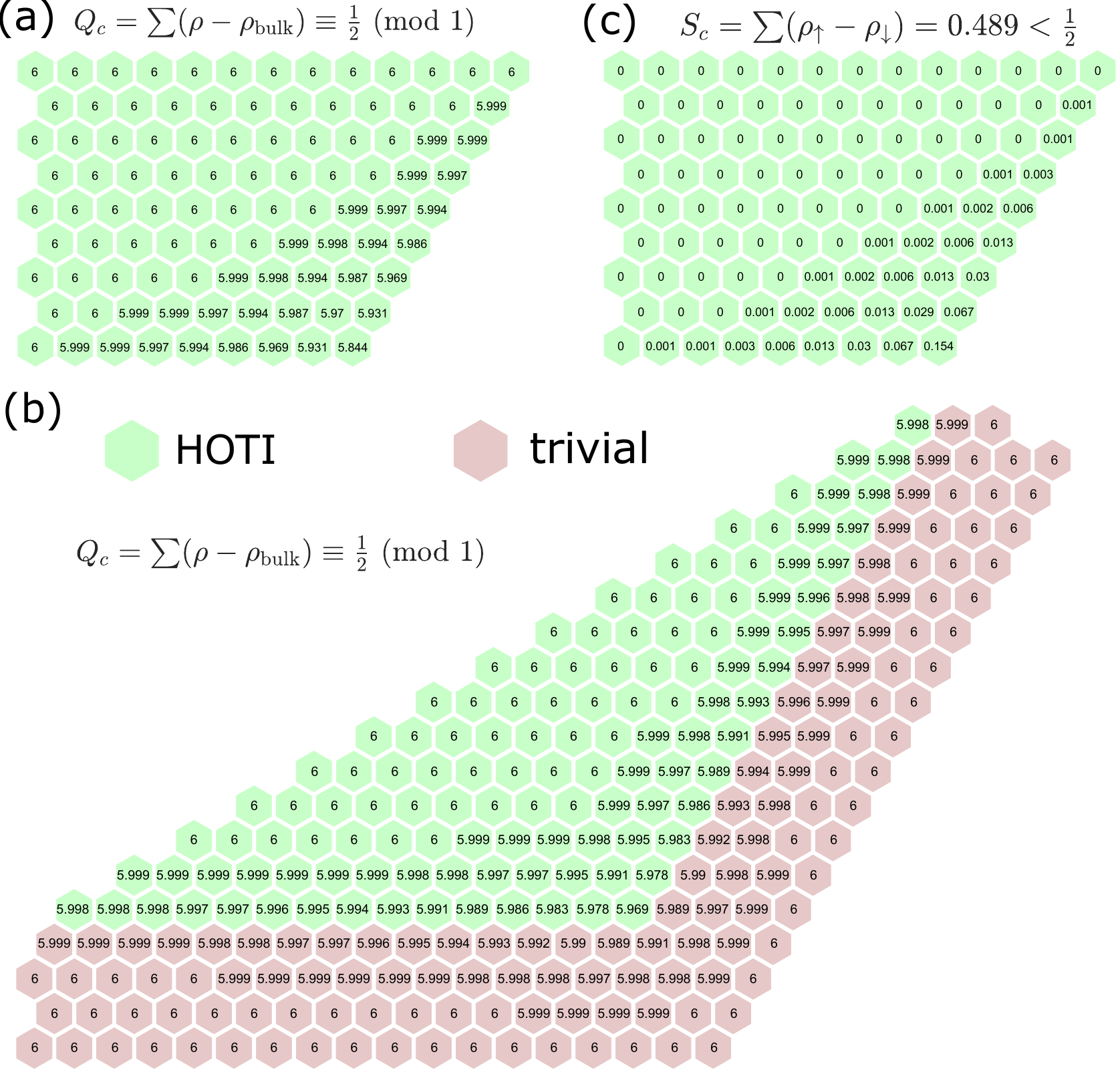}
    \caption{Accumulated LDOS $\rho$ pattern around an open or embedded corner. (a) HOTI phase ($t_{\text{in}}^{\uparrow}=-0.8$,$t_{\text{out}}^{\uparrow}=-1.2$,$t_{\text{in}}^{\downarrow}=-1.2$,$t_{\text{out}}^{\downarrow}=-0.8$) with a finite SOC $\lambda_{\text{SOC}}=0.1$. (b) HOTI domain (green hexagons) embedded in trivial domain (red hexagons); the trivial domain is set without SOC. (c) Accumulated net spin $\rho_{\uparrow}-\rho_{\downarrow}$ pattern in the HOTI phase considered in (a); the corner spin is not well-quantized as discussed in the main text.}
    \label{fig:LDOS}
\end{figure}

We emphasize that $Q_c$, $Q_c^{\uparrow}$, $Q_c^{\downarrow}$, and $S_c$ are (when properly quantized) bulk topological invariants even though these quantities are often treated interchangeably as boundary observables due to well-established bulk boundary correspondence (BBC)~\cite{benalcazar2017quantized,benalcazar2019quantization, noh2018topological}. To be specific, as a result of $Q_c=\frac{1}{2}$ defined in the bulk, a terminated $120^{\circ}$-angled corner would possess a corner-localized $\frac{1}{2}$ deficit of local density of states (LDOS) accumulated for all states below the mid-bandgap energy (here, zero energy), calculated according to:
\begin{equation}
    \rho_\uparrow \left( \mathbf{R} \right) = \sum_{E < 0} \sum_{i=1}^{6} \left| \psi_n \left( \mathbf{R},i,\uparrow \right)  \right| ^2,
\end{equation}
where the summation is done on all negative energy states excluding the zero energy mode that is at small negative energy due to finite size effects, $\psi_n$ is the eigenstate for the open boundary system, $\mathbf{R}$ is the position of the unit cell and $i$ is the sublattice index. Thus $\psi_n \left( \mathbf{R},i,\uparrow / \downarrow \right)$ is   the wavefunction value at position $\mathbf{R}$ sublattice $i$ with spin up or down. Fig.~\ref{fig:LDOS}(a) shows the accumulated LDOS $\rho$ pattern around an open corner in the HOTI phase ($Q_c=\frac{1}{2}$); we obtain the corner charge of half by integrating the deviation of accumulated LDOS from the value deep inside the bulk (i.e. the number of bands below the bandgap; $\rho_{\rm bulk}=6$ in this case). Since this is a topological property, this corner charge of $\frac{1}{2}$ doesn't change for the case of embedding the topological domain inside surrounding trivial domain, see Fig.~\ref{fig:LDOS}(b). Embedding into the trivial domain is not necessary for the tight-binding model, but it is necessary for a photonic emulation system, as a true open termination with vacuum is difficult to realize in our photonic platform. The embedding naturally decreases the size of the band gap between the edge states, thereby weakening the localization of the corner state along the edges.

In Fig.~\ref{fig:LDOS}(c), we also demonstrate the accumulated net spin texture $\rho_{\uparrow}-\rho_{\downarrow}$ around an open corner. The LDOS-based calculation of spin corner shown in Fig.~2(b) of the main text was obtained by $\langle S_z \rangle=\sum (\rho_{\uparrow}-\rho_{\downarrow})$. As discussed in the main text, this spin corner is not well quantized in the HOTI phase in the presence of SOC. However, at a reasonable strength of SOC $\lambda_{\text{SOC}}=0.1$, $\langle S_z \rangle$ retains a value very close to the bulk quantity $S_c $ and both are close to 0.5 featuring a moderately (not perfectly) robust preservation of corner-localized spin.

\section{Bulk boundary correspondence} \label{sec:BBC}
In the main text, we briefly argued that $Q_c$, $Q_c^{\uparrow}$ and $Q_c^{\downarrow}$ can be used to predict the number of corner states at the interface of two different domains. Here, we elaborate on the details of these crucial bulk boundary correspondences (BBC) in our spin-dependent kekule lattice model in the presence of spin-orbit coupling. For the spinless kekule lattice, it has been well-established that the topological origin of a corner state can be explained as a direct result of $Q_c=\frac{1}{2}$ and the chiral symmetry $S=\text{diag}(1,-1,1,-1,1,-1)$~\cite{noh2018topological,jung2020nanopolaritonic}. $Q_c=\frac{1}{2}$ gives rise to half corner-localized deficit of LDOS accumulated for the bands below zero energy, and the chiral symmetry ensures that the same deficit of half from the bands above zero energy. The two halves or, equally, whole of corner-localized LDOS deficit should be compensated by a state localized around a corner, and this corner state must be pinned at zero energy again due to the chiral symmetry. This line of arguments effectively proves BBC in Kekulé lattices.

In our spinful system, the simple extension of the previous chiral symmetry $S\otimes I_{2\times2}$ does not satisfy the condition of a chiral symmetry $H(S\otimes I_{2\times2})=-(S\otimes I_{2\times2})H$ with a finite $\lambda_{\text{SOC}}$; here, $H$ is from Eq.~(S1). In fact, a spin-dependent extension $S_{\text{SOC}}=S\otimes \sigma_{z}$ satisfies the given condition $HS_{\text{SOC}}=-S_{\text{SOC}}H$. Thus, our spinful system in the presence of spin-orbit coupling possesses the chiral symmetry given as $S_{\text{SOC}}$. Note that this new chiral symmetry operator, when applied to an eigenvector, still preserves the spin contents of the eigenvector; $\left<{\psi} \right| s_z \left| {\psi} \right> = \left<{\psi'}\right| s_z \left| {\psi'}\right>$ for $\left| {\psi'}\right>=S_{\text{SOC}}\left|{\psi}\right>$. This spin-preservation property is very crucial, as it allows us to apply the BBC of kekule lattices to each spin subspace.

Now that we have properly defined the spin-preserving chiral symmetry for our spinful Hamiltonian, let's establish BBCs for the following cases:

\subsection{Topological phase (for generic $\lambda_{\text{SOC}}$) with an open corner or embedded corner interfacing trivial phase}

With $\lambda_{\text{SOC}}=0$, the system is a simply decoupled stack of two topological kekule lattices of each spin species. Therefore, we have $Q_c^{\uparrow}=\frac{1}{2}$ and $Q_c^{\downarrow}=\frac{1}{2}$. Suppose that we now turn on the spin-orbit-coupling $\lambda_{\text{SOC}}>0$. Recall that we have established in the previous section that our spin-orbit coupling only induces couplings between the same orbital (dipolar or quadrupolor) states. Thus, there cannot be any coupling between the modes below zero energy and the modes above zero energy in topological phase, because all quadrupole(dipole)-like states of both spin species are below(above) zero energy. As a result, $Q_c^{\uparrow}=\frac{1}{2}$ and $Q_c^{\downarrow}=\frac{1}{2}$ are conserved even under a finite $\lambda_{\text{SOC}}>0$, since the exchange of spin expectation values is being done within the bands below zero energy.

Therefore, with and without spin-orbit-coupling, we always have well-quantized $Q_c^{\uparrow}=\frac{1}{2}$ and $Q_c^{\downarrow}=\frac{1}{2}$, and each of these invariants, along with the spin-preserving chiral symmetry $S_{\text{SOC}}$, predicts a whole deficit of corner-localized accumulated LDOS, which gives rise to total two corner states. There are two options for the energy of these corner states: They can either both be at zero energy, each being a chiral partner of itself, or they can be at $\pm E$ as chiral partners of each other. For the perturbations we consider the corner states are always chiral partners of themselves and are therefore at zero energy.

\subsection{HOTI phase (for generic $\lambda_{\text{SOC}}$) with an open corner or embedded corner interfacing trivial phase}

$Q_c=\frac{1}{2}$ and the chiral symmetry $S_{\text{SOC}}$ ensures a corner state at zero energy.

\subsection{HOTI phase (for generic $\lambda_{\text{SOC}}$) with an embedded corner interfacing topological phase}

$Q_c=\frac{1}{2}$ and the chiral symmetry $S_{\text{SOC}}$ ensures a corner state at zero energy ($Q_c=0$ in topological phase).

\subsection{HOTI with $Q_c^\uparrow = 1/2$, $Q_c^\downarrow = 0$ interfaced with HOTI with  $Q_c^\uparrow = 0$, $Q_c^\downarrow = 1/2$ (for vanishing $\lambda_{\text{SOC}}=0$)}

With vanishing spin-orbit-coupling $\lambda_{\text{SOC}}=0$, $Q_c^{\uparrow}$ and $Q_c^{\downarrow}$ are well-defined even in the HOTI phase; $Q_c^{\uparrow}=\frac{1}{2}$ and $Q_c^{\downarrow}=0$ for HOTI I, and $Q_c^{\uparrow}=0$ and $Q_c^{\downarrow}=\frac{1}{2}$ for HOTI II. Thus, for each spin subspace, two domains are topologically distinct with $Q_c$ difference of half, and therefore we obtain a corner state at zero energy.

We note that the BBC for the case of HOTI I interfacing with HOTI II is not  established for a finite spin-orbit coupling. We empirically find that the spin corner $S_c$, even though not perfectly quantized, is nearly conserved to be $+\frac{1}{2}$ for HOTI I and $-\frac{1}{2}$ for HOTI II, thus providing an effective distinction between these two phases for small values of SOC.

\section{Topological indices in QSH phase ($\Delta_\uparrow=\Delta_\downarrow$; $|\lambda_{\text{SOC}}|>|\Delta_\uparrow|$)}

When two spin subspaces are degenerate ($t^\uparrow_{\text{in/out}}=t^\downarrow_{\text{in/out}}\equiv t_\text{in/out}$; thus, $\Delta_\uparrow=\Delta_\downarrow\equiv\Delta$ and $H_\uparrow=H_\downarrow\equiv H_0$), by changing the spin basis $s_x\rightarrow s_z$, the system hamiltonian can be made to be block-diagonal, where one block (spin up) is $H_0+H_{\text{SOC}}$ and the other block (spin down) is $H_0-H_{\text{SOC}}$. Basically, in this basis, we have a more conventional form of Kane-Mele spin-orbit coupling where it induces an effective magnetic field to each of two decoupled spin subspaces with opposite signs. In the chern-insulator perspective, we get $C_{s_x=\uparrow}=C_{s_x=\downarrow}=0$ when $|\lambda_{\text{SOC}}|<|\Delta_0|$ and $C_{s_x=\uparrow}=+1$ and $C_{s_x=\downarrow}=-1$ when $|\lambda_{\text{SOC}}|>|\Delta_0|$ (here, the chern number is the composite chern number of the bands below the bandgap). Thus, the QSH phase $|\lambda_{\text{SOC}}|>|\Delta_0|$ carries the spin-chern number of $C_{s_x=\uparrow}-C_{s_x=\downarrow}=2$.

\begin{figure}
    \includegraphics[width=0.9\columnwidth]{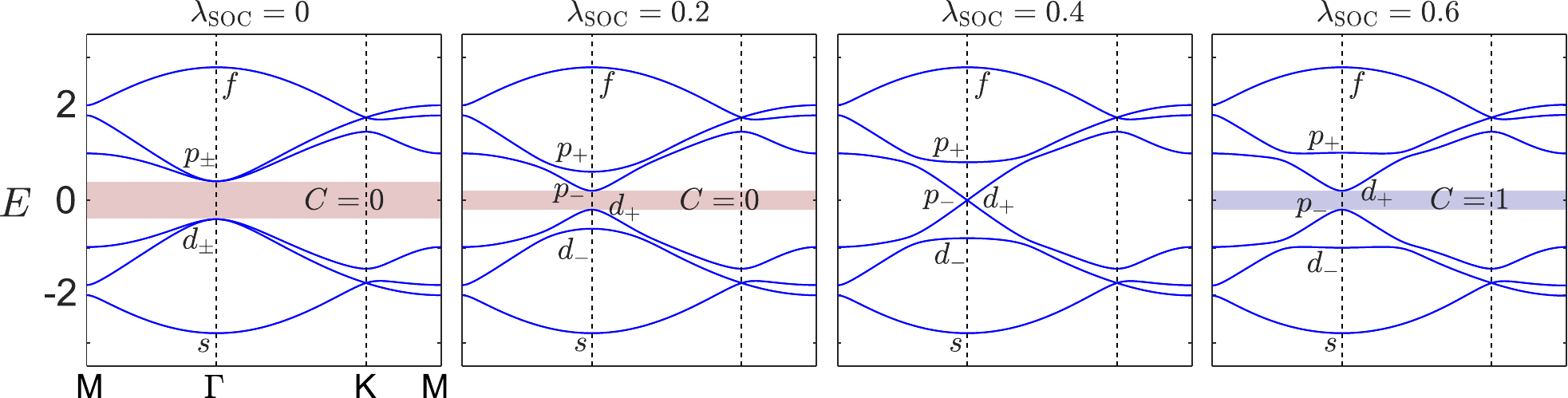}
    \caption{Band structures of a single $s_x=\uparrow$ spin subspace when the system is block-diagonalizable under a spin-degeneracy condition $t^\uparrow_{\text{in}}=t^\downarrow_{\text{in}}\equiv 0.8$, $t^\uparrow_{\text{out}}=t^\downarrow_{\text{out}}\equiv 1.2$; the strength of magnetic flux $\lambda_{\text{SOC}}$ is varied across the phase transition $\lambda_{\text{SOC}}=\Delta$ ($\Delta=0.4$ in this case).}
    \label{fig.S?.}
\end{figure}

In the corner charge index perspective, we get $Q_c^{\uparrow\downarrow}=\frac{1}{4}$, $Q_c=Q_c^{\uparrow}+Q_c^{\downarrow}=\frac{1}{2}$, and $S_c=Q_c^{\uparrow}-Q_c^{\downarrow}=0$. The spin-projected corner charges $Q_c^{\uparrow\downarrow}$ are well-quantized as $\frac{1}{4}$ in any spin-projection basis for the following reasons. First, let's consider $s_x$ basis where the system is favorably block-diagonalizable. Also, let's consider the $s_x=\uparrow$ subspace only. Across the phase boundary $|\lambda_{\text{SOC}}|=|\Delta|$ between the spin-HOTI phase and QSH phase, $\#\Gamma_{s_x=\uparrow}$ changes from $3$ to $2$, due to the band-inversion between $d_+$ and $p_-$ orbitals, see Fig.~\ref{fig.S?.}. In the meantime, $\# M_{s_x=\uparrow}$ stays as $1$. Thus, we observe the transition from $Q_c^{\uparrow}=\frac{\#\Gamma_{s_x=\uparrow}-\# M_{s_x=\uparrow}}{4}=\frac{2}{4}$ to $\frac{1}{4}$. For the $s_x=\downarrow$ subspace, the same transition occurs due to the band-inversion between $d_-$ and $p_+$ orbitals. The same principle applies for the phase transition between the trivial phase and QSH phase, resulting the change of $Q_c^{\uparrow\downarrow}=0$ to $\frac{1}{4}$.

\section{Parameters for tight binding models and photonic crystal designs}
Since our designs includes many parameters as shown in Fig.~3(a) of the main text, and not wishing to trouble the general reader, we include the design parameters of all phases plus SOC in the following tables, including what figure and subfigure they appear at, as can be seen in Tables \ref{table:TB} and \ref{table:PhC}. For the TB models the transition amplitudes between domain walls were set to be $\tilde{t}_\text{in}^{\uparrow,\downarrow} = (t_\text{in, outer domain}^{\uparrow,\downarrow}+t_\text{in, inner domain}^{\uparrow,\downarrow})/2 + \alpha(t_\text{in, outer domain}^{\uparrow,\downarrow}-t_\text{in inner domain}^{\uparrow,\downarrow})/2  $, where $\alpha$ is a real scalar representing the edge roughening. For $\alpha = 1$ there is no edge roughening and for $\alpha > 1$ the amplitudes between domain walls take some value between the amplitudes of the inner and outer domains. We've used $\alpha = 2$ in all our TB calculations. For the photonic first principle simulations the edge roughening used was setting the cylinders separating the domains to either $1.4\langle r_1 \rangle $ or $0.7\langle r_2 \rangle$ where $ \langle r_1 \rangle, \langle r_2 \rangle$ are the respective average radii of $r_1$ and $r_2 $ of the inner and outer domains. The air gaps between said cylinders and the top and bottom plates also took the average values of the air gaps of both domains.

\begin{table}[h!]
\centering
    \begin{tabular}{||c|c|c|c|c|c|c||}
    \hline
    Phase & Figures & $t_\text{out}^\uparrow$ & $t_\text{in}^\uparrow$ & $t_\text{out}^\downarrow$ & $t_\text{in}^\downarrow$ &$\lambda_\text{SOC}$ \\
    \hline \hline
    HOTI & Fig.2(a) & 0.7 & 1.3 & 1.2 & 0.8 & 0.8 \\
    \hline
    Spin-HOTI & Fig.2(b) & 1.3 & 0.7 & 1.2 & 0.8 & 0.4 \\
    \hline    
    Trivial & Fig.2(a),(b) & 0.7 & 1.3 & 0.8 & 1.2 & 0 \\
    \hline
    Spin-HOTI & Fig.4(a) & 1.1 & 0.8 & 1.4 & 0.8 & 0 (\text{solid lines}) 0.42 (\text{dashed lines})  \\
    \hline
    HOTI & Fig.4(b) & 0.8 & 1.1 & 1.4 & 0.8 & 0 (\text{solid lines}) 0.42 (\text{dashed lines})  \\
    \hline
    \end{tabular}
    \caption{Parameters for tight binding models.}
    \label{table:TB}
\end{table}

\begin{table}[h!]
\centering
    \begin{tabular}{||c|c|c|c|c|c|c|c|c|c|c||}
         \hline
         Phase & Figures & $h$ & $h_\text{mid}$ & $r_1$ & $r_1^\text{mid}$ & $r_2$ & $r_2^\text{mid}$ & $g_\text{top}$ & $g_\text{bot}$ \\
         \hline \hline
         HOTI & Fig.3(b) & a & 0 & 0.24a & 0.24a & 0.136a & 0.136a & 0.045a & 0.045a \\
         \hline
          Spin-HOTI & Fig.3(c), Fig.S8(a) & a & 0.2a & 0.22a & 0.08a & 0.11a & 0.21a & 0.02a & 0.02a\\
         \hline
         Spin-HOTI & Fig.4(e) & a & 0.2a & 0.22a & 0.08a & 0.11a & 0.21a & 0.02a & 0.04a\\
         \hline
         Spin-HOTI and HOTI transition & Fig.4(c), Fig.S8(b) & a & 0.2a & 0.22a & 0.08a & 0.11a & 0.21a & 0.02a & 0.075a\\
         \hline
         Spin-HOTI & Fig.S8(c) & a & 0.2a & 0.22a & 0.08a & 0.11a & 0.21a & 0.02a & 0.12a\\
         \hline
         HOTI & Fig.4(d) & a & 0 & 0.24a & 0.24a & 0.136a & 0.136a & 0.045a & 0.12a \\
         \hline
         Trivial & Fig.4(f) & a & 0.28a & 0.13a & 0.25a & 0.17a & 0.1a & 0.045a & 0.045a\\
         \hline 
         Trivial & Fig.4(e) & a & 0.06a & 0.08a & 0.236a & 0.17a & 0.1a & 0.045a & 0.045a\\
         \hline
         QSH & Fig.S6(b) & a & 0.2a &  0.22a & 0.04a & 0.132a & 0.22a & 0.015a & 0.1a\\
         \hline
         Trivial & Fig.S6(b) & a & 0.06a & 0.11a & 0.263a & 0.175a & 0.1a & 0.045a & 0.045a\\
         \hline
    \end{tabular}
    \caption{Parameters for photonic crystal designs.}
    \label{table:PhC}
\end{table}

\section{QSH phase and edge states}
\begin{figure}
    \includegraphics[width=0.8\columnwidth]{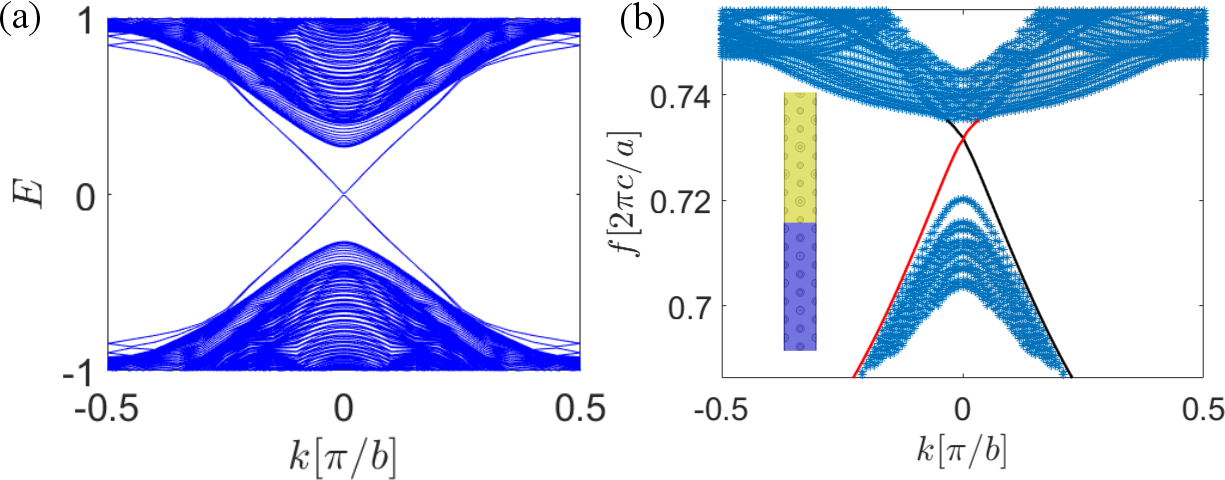}
    \caption{QSH phase. \textbf{(a)} TB model dispersion for $t_\text{out}^{\uparrow,\downarrow} = 1.2$, $t_\text{in}^{\uparrow,\downarrow} = 0.8$ and $\lambda_\text{SOC} = 1$ for the QSH domain and $t_\text{out}^{\uparrow,\downarrow} = 0.8$, $t_\text{in}^{\uparrow,\downarrow} = 1.2$ and $\lambda_\text{SOC} = 0$ for the trivial domain, with $\alpha = 2$ for cross domain hopping amplitudes. \textbf{(b)} Photonic supercell simulation dispersion results. Bulk modes in blue and edge modes in red and black. Inset: QSH phase in yellow interfacing a trivial phase in purple. See Table II for parameters.}
    \label{fig:QSH}
\end{figure}

As mentioned in the main text, when the TB model is tuned such that $\Delta_\uparrow = \Delta_\downarrow$ and $\lambda_\text{SOC} > \lambda_\text{th}$ a QSH phase is obtained and the corner states become delocalized. This was confirmed in a TB and photonic supercell simulation as shown in Fig.~\ref{fig:QSH}.

\begin{figure}
    \includegraphics[width=0.8\columnwidth]{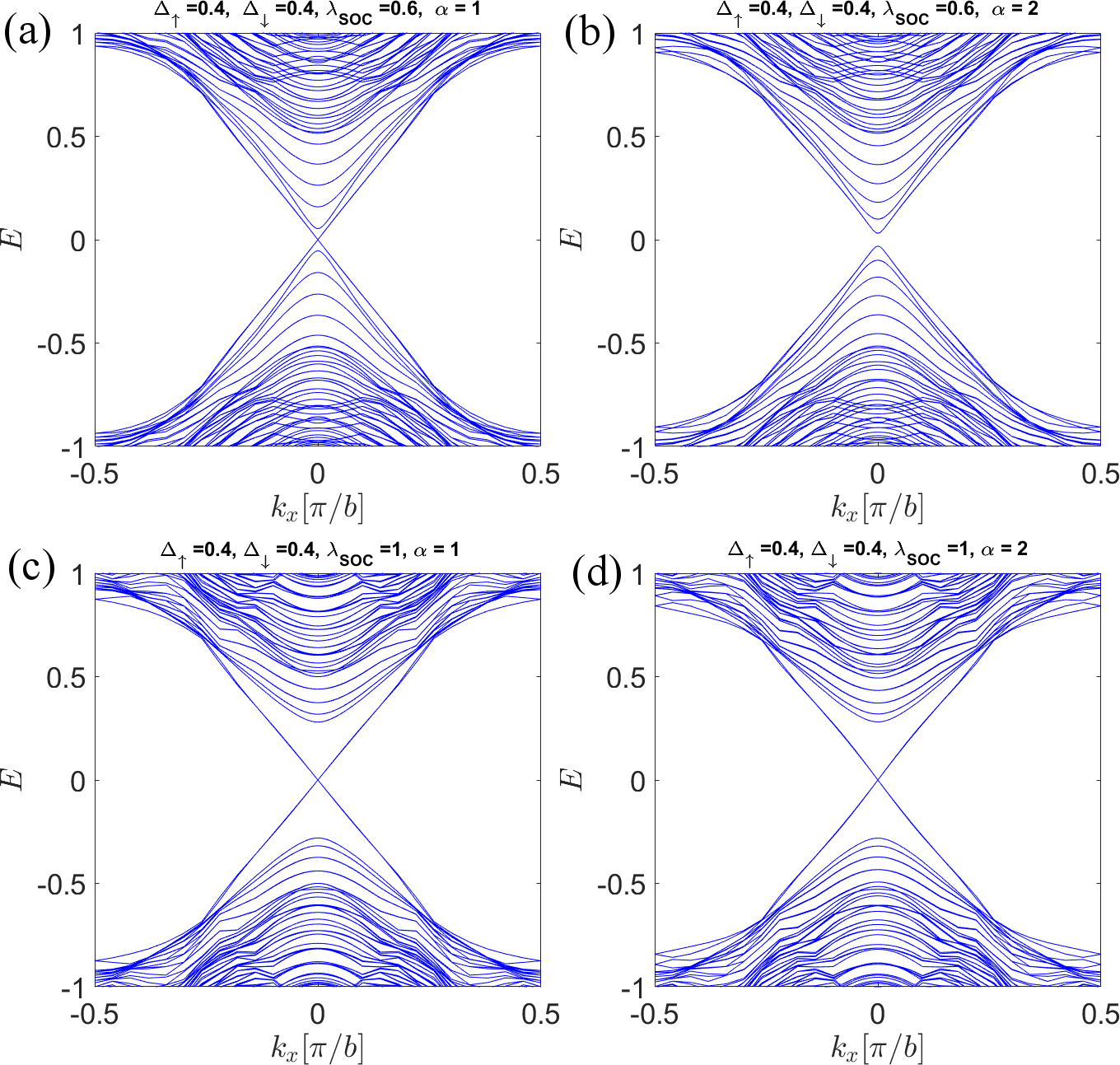}
    \caption{Effect of edge roughening on QSH edge states. \textbf{(a)} QSH with small bulk band gap and no roughening. \textbf{(b)} same as (a) but with edge roughening. \textbf{(c)} QSH phase with large bulk band gap and no edge roughening. \textbf{(d)} same as (c) but with edge roughening.}
    \label{fig:QSH_roughening}
\end{figure}

Additionally we investigated the effect of the edge roughening, i.e. changing the transition amplitudes along the domain wall between the QSH and trivial domains on the edge states. Edge roughening can gap the QSH edge states if the band gap is small enough, as can be seen in Fig.~\ref{fig:QSH_roughening}(a),(b). However, deep inside the QSH phase where the bulk band gap is large, edge roughening has no effect on the edge states and does not gap them, as can be seen in Fig.~\ref{fig:QSH_roughening}(c),(d).

\section{Photonic transition from spin-PHOTI to PHOTI}
\begin{figure}
    \includegraphics[width=0.9\columnwidth]{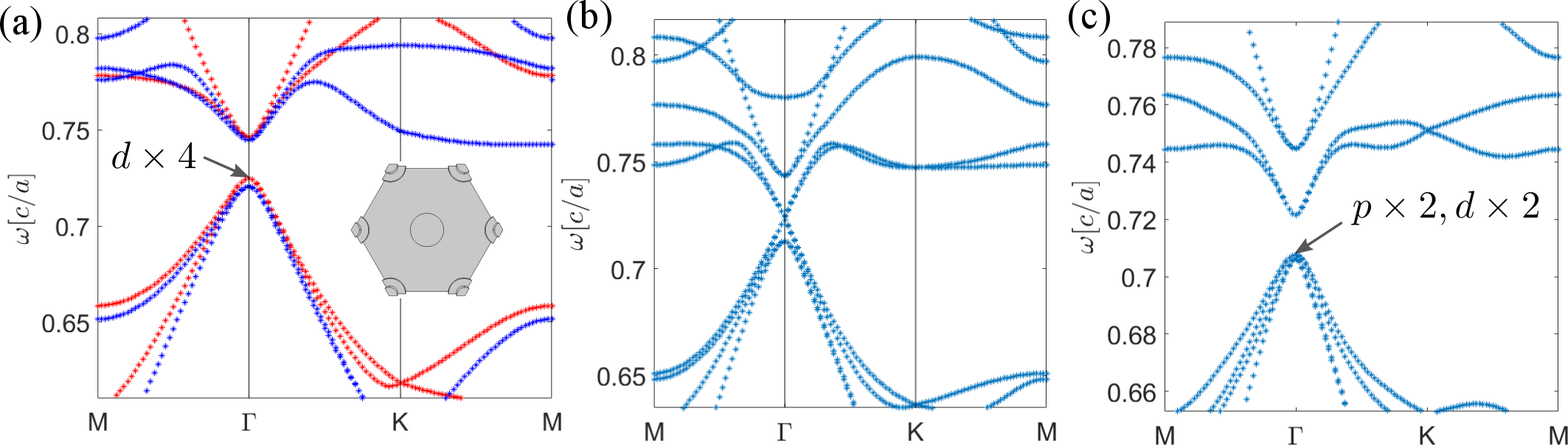}
    \caption{Photonic transition from spin-HOTI phase to HOTI phase. \textbf{(a)} Same as Fig.~3(c) from the main text showing HOTI phase with mid-plane mirror symmetry. All bands below the band gap are $d$ orbitals. \textbf{(b)}. Same as Fig.~4(e) from the main text showing tuning of mid-plane mirror symmetry breaking to close the band gap. \textbf{(c)} Further increase of the mid-plane mirror symmetry breaking causes the band gap to reopen with two $d$ and two $p$ orbital bands, indicating HOTI phase.}
    \label{fig:phaseTransition}
\end{figure}

Here we show that increasing $\Delta g$ can cause a phase transition from the spin-HOTI phase to the HOTI phase. We do this by taking the PhC design of Fig.~3(c) and Fig.~4(e) and further increasing $\Delta g$ to push the PhC to the HOTI phase. As can be seen in Fig.~\ref{fig:phaseTransition}(a), the PhC is in a spin-HOTI phase since all bands below the band gap are $d$ orbitals. An introduction of effective SOC by $\Delta g \neq 0$ closes the band gap in Fig.~\ref{fig:phaseTransition}(b), and a further increase of $\Delta g$ reopens the band gap, only now two of the bands below the band gap are $p$ orbitals. We note that the $p$ and $d$ orbital band gaps are not centered at the same frequency. This can be compensated by changing other geometric parameters of the design, but for the sake of the example we did not wish to change any parameters besides $g_\text{bot}$.

\FloatBarrier

\end{document}